\ifx\HCode\Undef        
  \DocumentMetadata{
    lang        = en,   
    pdfstandard = ua-2, 
    tagging     = off    
  }
\else                   
  \DocumentMetadata{}   
\fi                     

\documentclass[sigconf]{acmart}
\usepackage{algorithm}
\usepackage{algpseudocode}
\usepackage{xspace}
\usepackage{float}

\AtBeginDocument{%
  }

\setcopyright{acmlicensed}
\copyrightyear{2018}
\acmYear{2018}
\acmDOI{XXXXXXX.XXXXXXX}
\acmConference[SIGACCESS]{Make sure to enter the correct conference title from your rights confirmation email}{September 25--28, 2026}{Porto, Portugal}
\acmISBN{978-1-4503-XXXX-X/2018/06}

\newcommand{\name}{Audo-Sight\xspace}

\begin{document}

\title{Audo-Sight: Assistive Conversational Perception through Collaborative Edge-Cloud Intelligence}

\author{Jacob Bradshaw}
\email{jacob.bradshaw@unt.edu}
\affiliation{%
  \institution{University of North Texas College of Engineering}
  \city{Denton}
  \state{Texas}
  \country{USA}
}

\author{Mohsen Riahi Alam}
\email{mohsen.riahialam@unt.edu}
\affiliation{%
  \institution{University of North Texas College of Engineering}
  \city{Denton}
  \state{Texas}
  \country{USA}
}

\author{Bhanuja Ainary}
\email{bhanujaainary@my.unt.edu}
\affiliation{%
  \institution{University of North Texas College of Engineering}
  \city{Denton}
  \state{Texas}
  \country{USA}
}

\author{Minseo Kim}
\email{minseo.kim@unt.edu}
\affiliation{%
  \institution{University of North Texas College of Engineering}
  \city{Denton}
  \state{Texas}
  \country{USA}
}

\author{Mohsen Amini Salehi}
\email{mohsen.amini@unt.edu}
\affiliation{%
  \institution{University of North Texas College of Engineering}
  \city{Denton}
  \state{Texas}
  \country{USA}
}

\renewcommand{\shortauthors}{Bradshaw et al.}

\begin{abstract}
Despite advances in assistive technologies, Blind and Low-Vision (BLV) individuals continue to face challenges in understanding their surroundings. Delivering concise, useful, and timely scene descriptions for ambient perception remains a long-standing problem in accessibility. Existing solutions often fail to identify user expectations for real-time and accessible responses. Moreover, for a given task, they either rely on cloud offloading, which imposes a significant delay, or edge-based AI, which often sacrifices accuracy. To address this, we present Audo-Sight, an AI-driven assistive system that spans across Edge-Cloud continuum and enables BLV individuals to perceive their surroundings through voice-based conversation. \name provides low-latency, accurate, and human-friendly responses through a novel mechanism that seamlessly fuses Edge and Cloud responses. The system also addresses challenges in catering to BLV users through response editing informed by BLV needs. \name orchestrates a set of AI models based on user query contextual analysis to infer intent and adjust for a variety of situations. In urgent cases where users require  fast responses, Audo-Sight leverages parallel Edge and Cloud pipelines and seamlessly combines responses through its Response Fusion Engine. Systematic evaluation shows that \name delivers speech output around 80\% faster for urgent tasks and generates complete responses approximately 50\% faster across all tasks compared to a commercial cloud-based solution---highlighting the need for customized AI-based solutions. Human evaluation of \name shows that it is the preferred choice over GPT-5 for 62\% of BLV participants with another 23\% stating both perform comparably. Speed and interruption evaluations demonstrate that in most situations, the system can seamlessly respond at a rapid pace to keep up with BLV expectations.
\end{abstract}

\ccsdesc[500]{Human-centered computing~Accessibility systems and tools}
\ccsdesc[300]{Computing methodologies~Natural language generation}
\ccsdesc[300]{Networks~Cloud computing}

\keywords{Multimodal Large Language Models (MLLMs), AI Models, Edge-Cloud continuum, Context-Aware AI.}
\maketitle

\section{Introduction}\label{INT}
\subsection{Overview}
Assistive technologies aim to help Blind and Low-vision (BLV) individuals gain ambient perception; thus, they can perform their daily activities more efficiently and, in turn, can improve their independence and social inclusion. Despite remarkable advances in AI (particularly LLMs) and computing systems in recent years, their responses are still not timely and accurate enough for BLV individuals in several situation; hence, effective ambient perception continues to be a hurdle for people with BLV \cite{Penuela2025,Bhagat_2023}.

Effective ambient perception relies on back-and-forth interaction between users and assistive technologies to promptly analyze surroundings during daily tasks. For example, consider a BLV user ordering food in a restaurant without time pressure: the user must locate the counter, explore the menu, place an order, and complete payment. In contrast, a BLV traveler running late at an airport faces a far more urgent scenario, requiring rapid navigation to the correct terminal, identification of the departure gate, and timely completion of boarding procedures. A powerful conversational system to completely understand the user situation and awareness of surroundings through different AI models and sensors are required in the assistive technologies. Recent advances in collaborative intelligence across the edge-cloud continuum offer promising opportunities to realize active ambient perception for people with BLV. These include the multi-agent AI systems for tasks such as face recognition and object detection\cite{Envision,SeeingAI}, the integration of multimodal large language models (MLLMs) for vision-language reasoning \cite{Cheema:videodes}, and the progress in wearable assistive technologies across edge-cloud \cite{waisberg2024meta,enviglass}. In the software domain, all of these systems utilize multiple AI agents, and their performance depends on the performance of each individual agent as well as how they are managed within the edge–cloud continuum infrastructure.

\begin{figure}[!t]
    \centering
\includegraphics[width=\columnwidth, alt={The cloud interacts with the edge device. The edge interfaces with smart glasses. The smart glasses pick up image and record audio, and can be used in environments such as the airport pictured in front of the glasses. The edge and cloud include Multi-modal LLMs (MLLMs), and transfer image and text information between them.}]
    {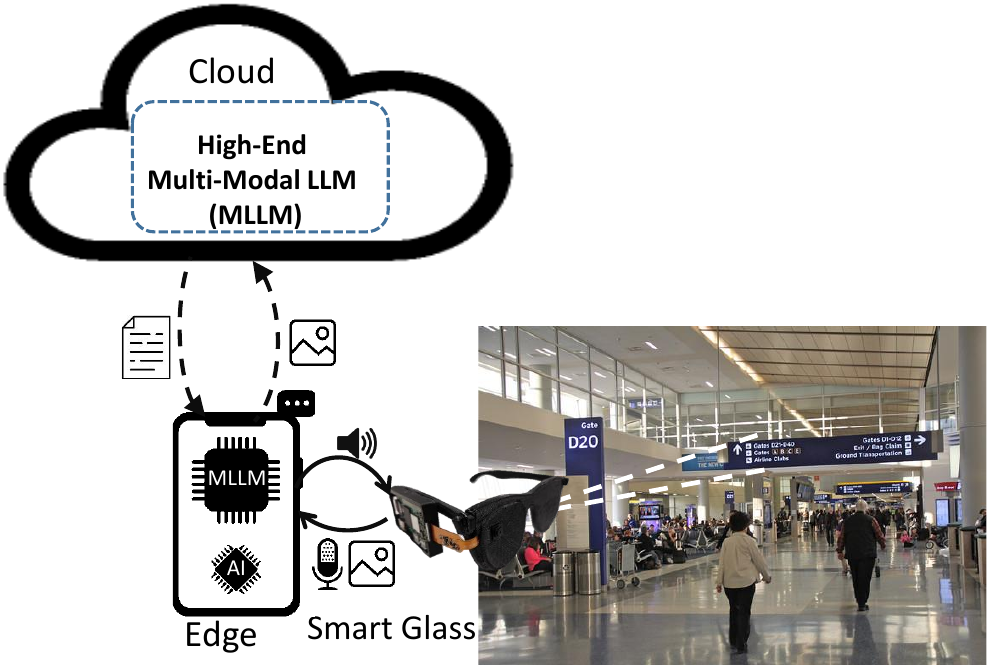}
    \vspace{-3mm}
    \caption{Overview of the \name framework that can provide conversational ambient perception for BLV individuals.}
    \label{fig:motive}
    \Description{The cloud interacts with the edge device. The edge interfaces with smart glasses. The smart glasses pick up image and record audio, and can be used in environments such as the airport pictured in front of the glasses. The edge and cloud include Multi-modal LLMs (MLLMs), and transfer image and text information between them.}
 \vspace{0mm}
\end{figure}

Accordingly, the goal of this study is to investigate the challenges inherent in integrating AI models and MLLMs on the edge–cloud continuum to enable a timely, accurate, context-aware, human-centric framework for BLV users. We develop \name, an AI-driven assistive framework for BLVs based on the smartglasses shown in Figure~\ref{fig:motive}. The smartglasses are paired with an AI-enabled mobile edge system and a cloud backend. User queries about the surrounding environment are conveyed to \name to be processed across edge-cloud and deliver context-aware and accurate responses while minimizing end-to-end latency. Within this framework, multiple AI models are deployed to address user intent, enabling the system to meet users’ needs across a range of everyday scenarios, particularly in time-sensitive or urgent situations. This framework also fuses and refines outputs from multiple MLLMs to make responses more aligned with BLV users’ requirements.

Efficient integration of multi-agent AI systems and MLLMs to assist BLV users presents several challenges, including selecting the appropriate AI models and tiers for deployment, determining how to invoke them based on user intent and the situation, and, when multiple AI models are required for a task, ensuring efficient coordination of the models and fusion of their outputs.

\subsection{Problem Definition}
Selecting appropriate AI models, LLMs, and their tiers along the edge–cloud continuum for deployment presents a key challenge, as these choices must be closely aligned with the specific needs of BLV users. For instance, both conversational intelligence and vision-based capabilities are essential components of such systems, and they can be deployed on edge devices, in the cloud, or across both, with each approach introducing distinct trade-offs in terms of latency, accuracy, availability, energy consumption, cloud performance jitter and cost \cite{Yi2025EcoAgentAE}.

Another key challenge is selecting the most appropriate AI models based on user intent and context. An assistive system must automatically identify and employ the most relevant AI models to generate an effective response. This process depends on accurate interpretation of the user query and an understanding of user intent. To enable this functionality, a lightweight and low-latency AI model must be deployed at the edge to infer the user’s intent and contextual information associated with each query and to route the request to one or more appropriate AI models or LLMs for processing.

Based on semi-structured interviews with BLV users of assistive technologies, we found that an effective conversational ambient perception system must account for \textit{urgent} situations. For example, consider the time-sensitive nature of responding to a traveler’s query about the “direction to the departure gate” at an airport, compared to a query about “the ingredients and calorie content of a menu item” at a restaurant. In the former case, the user prefers a concise and rapid response, whereas in the latter, the user expects a comprehensive and highly accurate response. Existing assistive tools often generate lengthy voice-based responses containing unnecessary details, which are difficult to track due to the sequential nature of voice \cite{Chang2024}, and can cause noticeable delays in user decisions that sometimes leading to serious consequences\cite{scirobotics2021}, such as accidents. To address this challenge, similar to the solution for the previous challenge, an edge-deployed AI model, can be integrated into assistive systems to distinguish, based on context, between queries that demand urgent, minimal, and fast responses and those that require more comprehensive answers.

The third challenge arises in certain scenarios, when multiple AI models across edge and cloud may generate responses to the same user query with varying levels of accuracy and latency. For instance, a smaller and faster MLLM deployed at the edge and a larger MLLM hosted on the cloud may be jointly invoked for tasks such as scene description. Their responses may be either consistent or contradictory. This necessitates the development of a mechanism to effectively fusing the outputs of these models, ensuring both low-latency responses and high-accuracy results for the user. Accordingly, this raises the question of how to seamlessly fuse the high-quality responses generated by cloud-based AI model with low-latency responses produced by edge-based AI model.

Interviews with BLV users further indicated that two design factors related to audio feedback can improve performance and user experience in the use of assistive devices. The first factor is how quickly the device initiates audio output. Minimizing latency is essential, as users expect audio feedback to begin as soon as possible to reduce perceived waiting time. A potential solution is for assistive technologies to initiate audio output as early as possible using a lightweight model that produces brief, less detailed information, and then continue with a more capable model that generates more accurate and comprehensive speech. However, implementing such a strategy is non-trivial. For instance, when outputs from the two models are consistent, the system must avoid redundant or repetitive content, whereas in cases of conflicting responses, it must ensure that the final output resolves discrepancies and preserves correctness.

The second is how the device can support user control of different speech rates in its audio feedback. Voice responses have a sequential nature, and BLV users typically have a much higher auditory perception than others. Therefore, it is very important for them to be able to increase the speech production rate, sometimes up to twice the normal speed or more. This effectively expands the bandwidth for information intake, allowing them to stay more quickly informed about their surrounding environment. Consequently, this feature should be considered in the design of assistive technologies, both during AI models selection and response generation and refinement, ensuring that speech synthesis is not compromised.


\subsection{Contributions}
In this study, we investigate these challenges and enable BLV users to effectively perceive their environment through \name and its conversational interactions. To mimic the perception abilities of sighted people, \name classifies incoming queries as either \textit{urgent} or non-urgent (i.e., \textit{normal}) and provides fast-track and normal-track pathways to support latency-driven and accuracy-driven interactions, respectively.

At the heart of \name, there is a ``Cognition Module'' that includes a set of expert AI models and MLLMs deployed within both edge and cloud, working in tandem to achieve a high accuracy and low latency perception reliably (robust against transient edge cloud disconnections).
Model selection was performed based on BLV user requirements in different environments and the available edge capacity. Cognition Module equipped with methods that dynamically employs the underlying AI models to balance the responsiveness of edge with the high-accuracy offered by high-end processing on the cloud, thereby, minimizing the latency without compromising accuracy for the urgent tasks, and maximizing accuracy for the non-urgent ones. 

For urgent tasks, we develop a mechanism that seamlessly fuses the rapid preliminary response from edge with a more accurate but slower response from the cloud to provide timely and reliable response to the user. This is very much like when a person responds quickly about an initial thought, while in the background of his/her mind prepares a more accurate response. Lastly, the generated responses must be adapted based on the perception capabilities of the users from the environment. As such, \name is equipped with a method that can assure blind-friendliness of the generated responses.

We have developed a prototype of the entire \name framework (including the smartglasses and the software platform across tiers), and curated a dataset of day-to-day images (taken by BLV users) and their pertinent urgent and non-urgent queries. These artifacts will be revealed upon acceptance of the paper. To conduct a comprehensive evaluation of \name, we conducted both system-level performance assessments and a human-study using BLV participants. 

In sum, the key contributions of this study are as follows:
\begin{itemize}
\item We developed \name as an assistive framework that uses edge–cloud AI solutions to provide timely, accurate, and context-aware perception through conversational interaction with BLV individuals. \name distinguishes between different queries and routes them accordingly to balance accuracy and latency requirements.

\item We propose ``Cognition Module'' as the reasoning engine of \name that can not only distinguish between urgent and normal queries but also select and utilize the most appropriate AI models across edge-cloud to accurately and swiftly meet user needs.

\item We develop a seamless edge–cloud response fusion mechanism that initiates audio output as early as possible by dynamically merging fast edge responses with more accurate cloud-generated responses. This mechanism enables both real-time and accurate assistance for users.
 
\item We examined \name with BLV human subjects and analyzed the accuracy, latency, and blind-Friendliness (accessibility) of the generated responses.

\end{itemize}

While \name is developed in the context of BLV users and assistive technology, many of its challenges and solutions are common across many other edge-cloud-AI systems and they can be adapted to be used for them. Specifically, this study addresses the broader challenges related to: The role of edge relative to cloud; Achieving accuracy and low latency in AI Fusion across edge and cloud; And transient disconnected operation of the edge.
The rest of this paper is organized as follows: Section II reviews the relevant literature, while Section III describes the proposed solution and methodology of \name. Section IV presents the experimental evaluation and BLV human study. Finally, Section V concludes the paper with a summary of the key findings.

\section{Related Work}
Nowadays, many assistive systems for BLV individuals leverage multiple AI models that work together across various platforms to address users’ needs \cite{NaviSense2025,WorldScribe2024}. As mentioned in the introduction, many challenges in this area arise from how to achieve more efficient integration of AI models and better coordination among them. Therefore, we first review prior work on the integration of AI models that address problems closely related to our own, with a focus on approaches that tackle similar challenges. Subsequently, to ensure that our proposed method aligns with the needs of BLV users, we examine domain-specific studies in assistive technologies, highlighting how these works have approached and addressed challenges in BLV-assistive system design.
Previous studies have emphasized the importance of selecting appropriate AI models for accessibility-focused applications. In particular, several works suggest that leveraging powerful cloud-based LLMs~\cite{BeMyAI,WorldScribe2024,Cheema:videodes,merchant:blvnav} can significantly enhance system performance. For instance, Cheema et al. \cite{Cheema:videodes} proposed an audio description system that directly prompts GPT-4V\cite{2023GPT4VisionSC} to generate detailed audio descriptions tailored for BLV users. NaviSense \cite{NaviSense2025} employs two powerful cloud-based models, an LLM for voice interaction, GPT-4o-mini \cite{openai2025_gpt4omini} and MoonDream 2B \cite{moondream2025} Vision Language Models (VLM) for open-vocabulary object detection. WorldScribe\cite{WorldScribe2024} generates rich, hierarchical descriptions (general overviews to fine-grained details) using cloud-based GPT-4V \cite{2023GPT4VisionSC} to support BLV scene understanding. {Compared to the cloud-based BLV assistance systems, we also use a cloud-based LLM, but our focus is on reducing response latency and minimizing waiting times in urgent situations via expert models and edge MLLM.}

Alongside using powerful cloud-based AI models, assistive technologies also utilize some smaller AI models deployed on edge devices. Almost all of them incorporate speech-to-text (STT) and text-to-speech (TTS) models to support conversational interaction with BLV users. WorldScribe \cite{WorldScribe2024} leverages YOLO-World \cite{cheng2024yolo} on the edge to perform real-time object detection, producing word-level labels to support immediate scene understanding. Several assistive technologies\cite{10303628}\cite{ML-BLV} also deploy small on-device AI models for tasks such as document recognition, basic OCR, currency recognition, and simple light or color detection. These capabilities support offline operation and ensure low-latency responses to meet the real-time needs of BLV users. {
For urgent generic queries, unlike existing edge-based AI systems (e.g., \cite{SeeingAI, Envision, TapTapSee}) that  target offline operation or low-latency inference, our approach leverages edge MLLM as part of a coordinated latency-reduction strategy that combines edge and cloud responses to reduce response time while maintaining cloud-level response quality.}

BLV users have diverse needs that require ensembling AI models to improve service quality and performance. However, using model ensembles introduces a routing challenge before inference: selecting the most suitable models for each query without running all models, to balance accuracy, latency, and efficiency. Existing approaches can be divided into training-free routers and training-required routers\cite{chen2024role}.
LLM routing \cite{ong2024routellm,ding2024hybrid,hu2024routerbench} uses a router to assign each query to the most suitable model, directing simple queries to smaller LLMs and complex ones to larger models to balance quality and cost. RouterDC \cite{chen2024routerdc} introduces a learned router that selects the most suitable LLM for each query. It uses two specialized learning techniques called contrastive losses to match queries with models and to capture query similarities. Results show that it outperforms existing routing approaches, with strong generalization to unseen data.

After the inference phase, if multiple AI models have been invoked, how their outputs are utilized becomes important. In some applications, a more accurate answer can be obtained by leveraging multiple responses from different models. This is achieved by sending a user query to multiple AI models and subsequently merging their outputs to produce a final response. In \cite{jiang2023llmbln}, a fusion module is proposed to combine multiple responses by leveraging their strengths and mitigating their weaknesses, formulated as a sequence-to-sequence task. Following this approach, the authors employ Flan-T5-XL \cite{flant5merge}, a 3B-parameter model, as the fusion component due to its strong performance and relatively compact size. {In this work, we evaluate this fusion-based approach against our proposed fusion method under the same settings.}

A staged response strategy is proposed in \cite{WorldScribe2024} to begin generating audio as quickly as possible by using a multi-model description pipeline that sends scene frames to three AI models with different latency and granularity levels. It streams outputs sequentially, first returning faster object-level labels, followed by short descriptions with spatial relationships, and finally detailed visual descriptions from the strongest model. The system does not fuse, merge or select outputs, but presents them in order as they become available to the user.
{Compared with the strategy proposed in \cite{WorldScribe2024}, our approach places greater emphasis on simultaneously minimizing response latency and maintaining high response quality, particularly in urgent scenarios. The two approaches differ in two key aspects: \textit{First}, our method selects models based on the urgency and context of the user's query, whereas WorldScribe \cite{WorldScribe2024} dynamically switches between models according to frame stability and user hand movements. While effective for its target application, this strategy cannot efficiently handle urgent queries. \textit{Second}, WorldScribe presents the outputs of multiple models directly to the user, which can result in redundant or inconsistent information. In contrast, our approach incorporates a fusion engine that identifies and removes duplicate content while resolving conflicts between model outputs during the generation of the final response. Moreover, when responses from smaller or weaker LLMs are incomplete or inaccurate, outputs from stronger LLMs can compensate for these limitations, producing more reliable and comprehensive responses.}

The authors in \cite{mahadevfuse} proposed Multitier AI fusion that enables combining results from concurrent execution across multiple tiers to achieve both low latency and high accuracy. Efficiency is achieved by running lightweight models on device, eliminating network overhead, while higher tiers improve accuracy using larger models and additional knowledge not available at the edge, including aggregated data from other devices and inferred or archived information that cannot be deployed to edge devices or restricted data. In latency-sensitive settings, immediate use of device outputs may be preferred over waiting for higher-tier results.
{Compared with Multitier AI Fusion \cite{mahadevfuse}, which employs a higher-tier model to aggregate information from lower-tier models, our approach relies on the cloud-based LLM as the primary source of accurate responses due to its superior reasoning capabilities, while leveraging edge-generated responses to reduce response latency.}

Overall, the contribution of our work is to offer an efficient integration and coordination method across multiple AI models to support urgency and low-latency response time simultaneously.

\section{\name: A Multi-Modal Smart Assistive Technology across Edge-Cloud}

\subsection{\name Architecture}\label{ARCH}

\begin{figure*}[!t]
    \centering
\includegraphics[width=0.99\linewidth, alt={A bird-eye view of the Audo-Sight architecture illustrates its primary components and data flow, beginning on the left with user input from a person wearing smart glasses. Data from the smart glasses and a microphone feeds into the Input Management module, which processes the "Image Input" and "Voice Input (query & prosody)" and directs both into a "Query-to-Image Mapper." The data then flows into the Cognition Module, passing through a "Query Router" and into a "Reasoning Engine" consisting of Expert AI-Models and Multi-modal LLMs. Following this, the processed information enters the Response Management module, moving through a "Response Refiner" before reaching a "Text to Speech" block. Finally, the text-to-speech output is directed to a speaker icon on the far right, representing the final audio response provided back to the user.}]
    {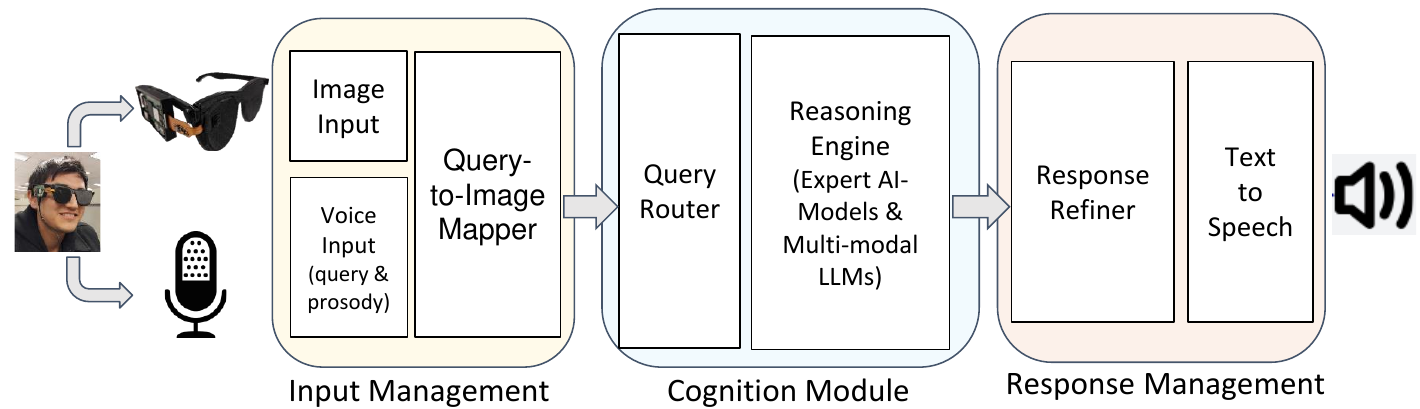}
    \caption{A bird-eye view of the \name architecture and its primary components}
    \label{FigArch}
    \Description{A bird-eye view of the Audo-Sight architecture illustrates its primary components and data flow, beginning on the left with user input from a person wearing smart glasses. Data from the smart glasses and a microphone feeds into the Input Management module, which processes the "Image Input" and "Voice Input (query & prosody)" and directs both into a "Query-to-Image Mapper." The data then flows into the Cognition Module, passing through a "Query Router" and into a "Reasoning Engine" consisting of Expert AI-Models and Multi-modal LLMs. Following this, the processed information enters the Response Management module, moving through a "Response Refiner" before reaching a "Text to Speech" block. Finally, the text-to-speech output is directed to a speaker icon on the far right, representing the final audio response provided back to the user.}
\end{figure*}

The bird-eye view of the \name architecture is shown in Figure~\ref{FigArch}.
The input device in \name is a custom-designed pair of smart-glasses that we developed to capture frames from the surrounding environment along with the user's voice query. 
Upon receiving the voice-based query, it is converted to text (denoted as $Q$) and paired with the pertinent frame (denoted as $f$) in the Query-to-Image mapper of the Input Management module. The urgency of the user's query can be identified either from the voice prosody or the text version of it. The $(Q,f)$ pair is then forwarded to the Cognition Module of \name.

The Cognition Module, one of the main contributions of this study, acts like the brain of \name. At its kernel, it contains the Reasoning Engine, which consists of a set of lightweight \textit{expert} (i.e., specialized) AI models (e.g., Optical Character Recognition (OCR), facial recognition, and object detection) to handle specific queries, and heavyweight MLLMs to handle more general queries. The MLLMs are specifically spread across edge and cloud for two reasons: (a) dealing with the resource and energy limitations of the edge; and (b) assuring availability (reliability) of the interaction even in the face of  network failures and cloud inaccessibility. Upon receiving $(Q,f)$ pair, the query routing mechanism first detects the urgency of the query and accordingly routes it to the appropriate subsystems with the best-matching AI model in the Reasoning Engine to handle the query based on the given visual context $f$.

Once a response is generated by the Reasoning Engine, it is further processed and refined by the Response Management module before speech synthesis. In analogy, this module is very much like the cognitive control of our brain that evaluates our thoughts before speaking them. The response refining process can include a wide range of actions. It can customize the raw response generated by the Reasoning Engine to become human-friendly and even blind-friendly via removing or altering the response to ensure effective communication with the user. Importantly, one aspect of Response Refiner, called Response Fusion Engine (described in~\ref{subsubsec:ResFusEng}), is in charge of seamlessly fusing responses generated by both edge and cloud tiers for the same query. This ensures the  communication of a fast response (generated by edge) and an accurate response (generated by cloud).

\subsection{Input Management Module}
The smartglasses used for \name are 3D printed frames containing a Raspberry Pi Zero 2 W with a camera. The smartglasses device is connected to the \name edge computing platform via WiFi or USB. When the user begins a query, triggered by an input button event, an image frame ($f$) is captured from the glasses video stream and transferred to the edge system via TCP. The frame, $f$, from the start of the interaction is analyzed along with the user's voice recording. The textual query, $Q$, is extracted from the voice input using the Whisper STT engine \cite{radford2022}. These two pieces of input data are paired together in the Query-to-Image Mapper component and sent to the Cognition Engine. Our evaluation focuses on the \name edge-cloud processing using pre-captured images and datasets rather than the smartglasses.

\subsection{Cognition Module}
\begin{figure*}[!t]
    \centering
    \includegraphics[width=0.99\linewidth, alt={The system architecture is divided into two main sections: a Cognition Module and a Response Management Module. The input Query ($Q$) enters the Urgency Detector within the Cognition Module, while an image Frame ($F$) feeds directly into the Generic AI Models block of the Reasoning Engine. The Urgency Detector processes the query, sending it along either a normal or urgent track into an AI Router, which then directs the flow into the Reasoning Engine as either an Expert or Generic Query. The Reasoning Engine consists of Expert AI Models (edge-based, featuring facial recognition, OCR, and object detection) to handle expert queries, and Generic AI Models (Cloud and Edge MLLMs) to handle generic queries alongside the frame input. The processed tracks then enter the Response Management Module. Here, the normal track for expert queries passes through a Blind-Friendly Response Editor—which also incorporates the original query ($Q$)—before entering the Text To Speech (TTS) block, whereas the urgent expert track flows directly to TTS. For generic queries, the normal track flows directly from the Cloud MLLM to TTS, while the urgent generic track routes from the Edge MLLM through a Response Fusion Engine before reaching TTS. Finally, the Text To Speech block consolidates all flows into a single audio output.}]
    {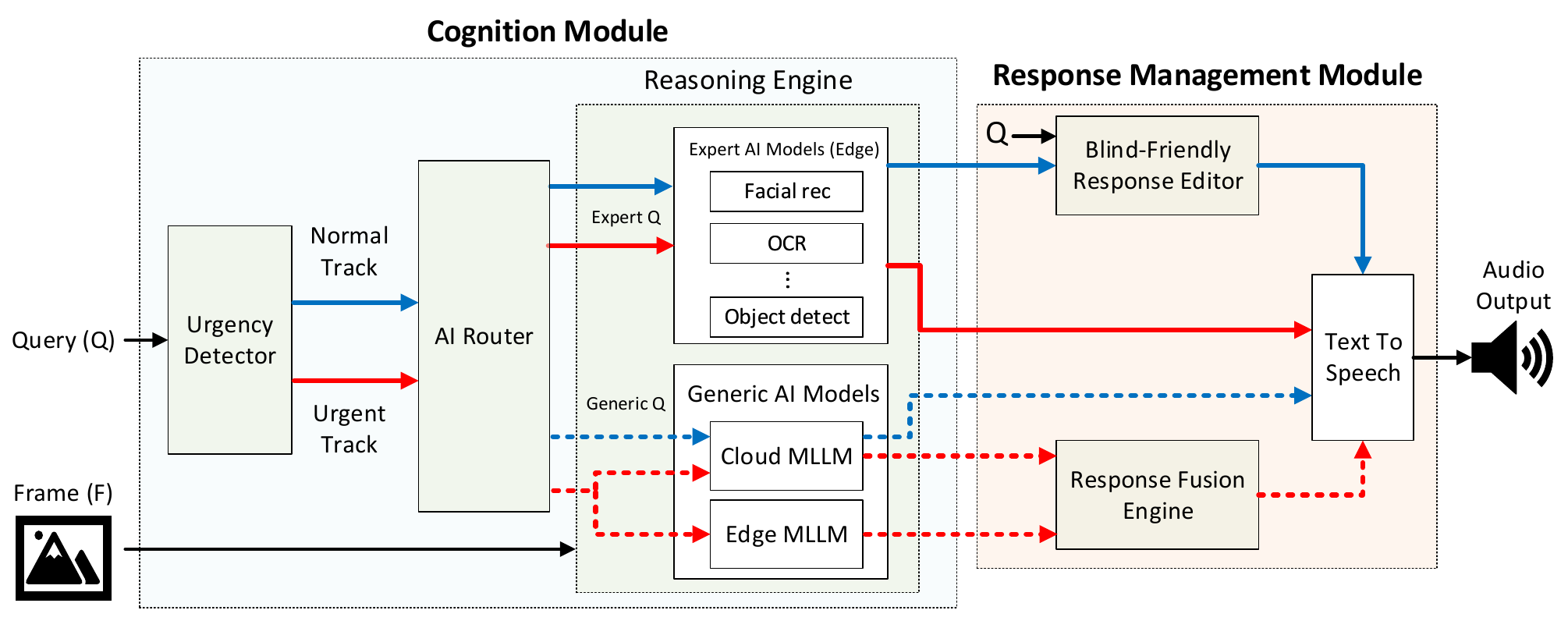}
    \caption{Internal mechanics of the Cognition and Response Management Modules of the \name platform}
    \label{fig:cog}
    \Description{The system architecture is divided into two main sections: a Cognition Module and a Response Management Module. The input Query ($Q$) enters the Urgency Detector within the Cognition Module, while an image Frame ($F$) feeds directly into the Generic AI Models block of the Reasoning Engine. The Urgency Detector processes the query, sending it along either a normal or urgent track into an AI Router, which then directs the flow into the Reasoning Engine as either an Expert or Generic Query. The Reasoning Engine consists of Expert AI Models (edge-based, featuring facial recognition, OCR, and object detection) to handle expert queries, and Generic AI Models (Cloud and Edge MLLMs) to handle generic queries alongside the frame input. The processed tracks then enter the Response Management Module. Here, the normal track for expert queries passes through a Blind-Friendly Response Editor—which also incorporates the original query ($Q$)—before entering the Text To Speech (TTS) block, whereas the urgent expert track flows directly to TTS. For generic queries, the normal track flows directly from the Cloud MLLM to TTS, while the urgent generic track routes from the Edge MLLM through a Response Fusion Engine before reaching TTS. Finally, the Text To Speech block consolidates all flows into a single audio output.}
\end{figure*}

The Cognition Module consists of three primary components, shown in Figure~\ref{fig:cog}. 
Upon receiving query $Q$, first, the edge-based \textit{Urgency Detector} determines whether a user requires a rapid response and accommodates this by activating the ``urgent track'' subsystem in the Reasoning Engine and Response Management that offer faster response times. Second, the edge-based \textit{AI Router} infers the intent underlying the user’s query, enabling the selection of task-appropriate AI models through routing. Finally, the \textit{Reasoning Engine} comprises two sub-parts, the Expert AI Models (on the edge) and the Generic AI Models (on both edge and cloud) that together provide access to a range of generic and specialized AI models to efficiently generate accurate responses. In the following sections, we provide a detailed explanation of the methods developed for each one of these components.

\subsubsection{Urgency Detector}
To craft an accurate urgency detection system from the human language (i.e., the user queries), we fine-tuned all-MiniLM-L6-v2~\cite{reimers2019sentencebert} (aka MiniLM) sentence-transformers model using the SetFit framework~\cite{tunstall2022}. We employed MiniLM because of its lightweight and edge-friendly design. We examined it against another popular model, called RoBERTa \cite{liu2019robertarobustlyoptimizedbert}, that is used for similar purposes. Our evaluations showed that the F1 scores of MiniLM were slightly more than RoBERTa (0.94 vs 0.93). However, MiniLM far outpaced RoBERTa in terms of speed, with MiniLM processing queries more than twice as fast on the GPU and 4 times as fast on the CPU, which proves its edge-friendliness. 

To further increase the speed of MiniLM, we employed SetFit to tune the model based on an urgency training dataset that we crafted to train the MiniLM model for urgent queries. With the tuning, we could substantially reduce the urgency detection overhead from 30 milliseconds to only 9 milliseconds. 
The training dataset was created using requests from VizWiz dataset \cite{Gurari_2018_CVPR} for the non-urgent (normal) class, and altered versions with urgent phrasing to create an urgent classification category. Although only the urgent class is used for classification, including a normal category helps differentiate normal requests from urgent ones due to the contrastive learning technique employed by SetFit. The model is trained offline, and the resulting urgency detection model is loaded once at runtime on the edge to evaluate every query. A threshold value is set for the urgency category, such that a query is considered urgent if its estimated probability for the urgent class exceeds this threshold. Based on the initial evaluations of the system, we chose to set the urgency threshold to 0.3 in the experiments. 

\subsubsection{AI Router} The AI Router conducts a secondary analysis of incoming queries following the urgency detector, in order to determine and route the $(Q,f)$ pair to the appropriate AI model within the Reasoning Engine. For decision-making, the AI Router employs a fine-tuned all-MiniLM-L6-v2 model; the same model is utilized in the Urgency Detector. We argue that although using transformer-based methods generally implies more overhead in the system design, however, it avoids developing and deploying multiple solutions (e.g., one for urgency detector and one for AI routing). Hence, it can simplify the system design and maintenance.

The MiniLM model in the AI Router classifies queries into three categories: Face Recognition, OCR, and Object Detection that we have so far implemented for the Reasoning Engine. The probability estimate for the highest-scoring category is compared against a predefined threshold value (we set it to 0.86 based on our initial evaluations), and the corresponding route to Expert AI models is selected if this threshold is exceeded. The queries that do not meet the threshold are classified as generic request and are routed to the Generic AI models in the Reasoning Engine.

In summary, as shown in Figure~\ref{fig:cog}, for query $Q$, Urgency Detector and AI Router work in tandem to route it through one of the following pipelines: Normal-track Expert, Urgent-track Expert, Normal-track Generic, and Urgent-track Generic.

\subsubsection{Reasoning Engine}
The Reasoning Engine comprises two sub-parts: the \textit{Expert AI Models} and the \textit{Generic AI Models}. The Expert AI Models are orders of magnitude faster than MLLMs and are responsible for generating response content from image analysis. In the Urgent-track, the generated response is directly delivered to the user; whereas, in the Normal-track, the response is sent to the Response Editor (in the Response Management module) for further processing and making it more blind-friendly. 

Within the Generic AI Models part, both the cloud-based MLLM GPT-5~\cite{openai_gpt5_2025} and the edge-based MLLM Gemma3:4B~\cite{geminiteam2025geminifamilyhighlycapable} are employed to process the assigned $(Q,f)$ pairs and handle general queries (i.e., those which cannot be handled by Expert AI Models). As there is no urgency associated with Normal-track Generic queries, they are only routed to the high-accuracy MLLM on the cloud. In addition to improving response accuracy, this preserves the edge battery lifetime and in case of a transient network failure, the user can resubmit the query (much like a best-effort request).
The queries in the Urgent-track, however, are routed to both the edge and cloud MLLMs to be processed more quickly and reliably (assuring a response in case of network unavailability). Because in this case two responses are generated for the same urgent query, there must be a subsequent method, called Response Fusion Engine in the Response Management Module, to seamlessly combine the high quality cloud response with the fast paced edge response.

\subsection{Response Management Module}
The Response Management serves as an interface between the Cognition Module and the user, refining the output responses from different AI models in each track category to suit BLV users. It delivers responses to users and includes Blind-Friendly Response Editor, Response Fusion Engine and a TTS component to synthesize voice output. BLV users typically prefer faster speech output rates relative to sighted users, so this module allows users to adjust the rate of speech production. This speed increment must not cause interruptions when the system is speaking and therefore requires careful consideration based on the AI models' token generation rates and capabilities. 

As illustrated in Figure~\ref{fig:cog}, the urgent responses generated by the Expert AI Models and normal responses generated by the Generic AI Models are transmitted directly to the TTS module. In contrast, normal responses produced by the Expert AI Models, together with the corresponding user queries, are first forwarded to the Blind-Friendly Response Editor for additional processing prior to TTS conversion. Furthermore, Urgent responses from the Generic AI Models are routed to the Response Fusion Engine, where supplementary processing is performed concurrently as the responses are being passed to the TTS module. The subsequent sections provide a detailed description of each of these components.

\subsubsection{Blind-Friendly Response Editor}
At times, the raw mixture of experts system outputs are not optimally user-friendly. This can be attributed to several causes: raw data wording problems, inaccuracies, or irrelevance. When considering the user query, the raw data wording may seem awkward or unresponsive to the user's request. For instance, in response to the query: ``Is there anyone I know here?" the raw facial recognition (an expert AI model) data is ``John Doe," but a better response would be ``Yes, John Doe is here." Similarly, inaccuracies such as a missing character from OCR can be accounted for by this system. Some information in the raw outputs can be irrelevant to the user's query, such as people listed in the object detection's response to ``What are these objects?", these responses undergo editing through Blind-Friendly Response Editor by being given to an edge-side LLM along with the user's original query. The LLM is prompted to answer the query using the results from the image analysis, and the output is streamed to the user. This component adds a notable latency overhead, so it is bypassed during urgent requests.
In the future, this subsystem can be extended to enhance accessibility in the systems' responses even further. During the system tests, we asked the BLV individuals how the system's responses could be improved, and some of their suggestions could be implemented the Blind-Friendly Response Editor subsystem, though updated to have generic responses routed through it as well. The specific feedback is described in the Experimental Study/Evaluation section~\ref{subsec:human}.

\subsubsection{Response Fusion Engine}
\label{subsubsec:ResFusEng}
The Response Fusion Engine works with the Cloud MLLM, Edge MLLM, and the TTS technology to provide a seamless transition from the low-latency Edge MLLM response to the more accurate, but slower cloud MLLM output. The lower accuracy Edge MLLM can make mistakes, and because the same request is evaluated by two independent models, this can introduce redundant or conflicting information. This component seamlessly fuses the two responses in real-time while self-correcting mistakes and avoiding detail repetition. This system is triggered during urgent, generic requests, and it leverages the simultaneous streaming of both the edge and cloud MLLMs. 

\begin{figure}
    \centering
    \includegraphics[width=1\linewidth, alt={A timeline diagram in image_a69b7c.png showing the parallel processing sequence of Edge and Cloud MLLMs. At t0 (Query Received), both models begin processing simultaneously. Edge Text-To-Speech (TTS) starts early at t1. At t2, the Cloud MLLM completes its processing, which abruptly terminates the ongoing Edge MLLM and its TTS (both marked with red Xs) and immediately triggers the Fusion Engine Processing. The Fusion Engine begins its own TTS output at t3, finishes generating at t4, and the final response completes at t5.}]{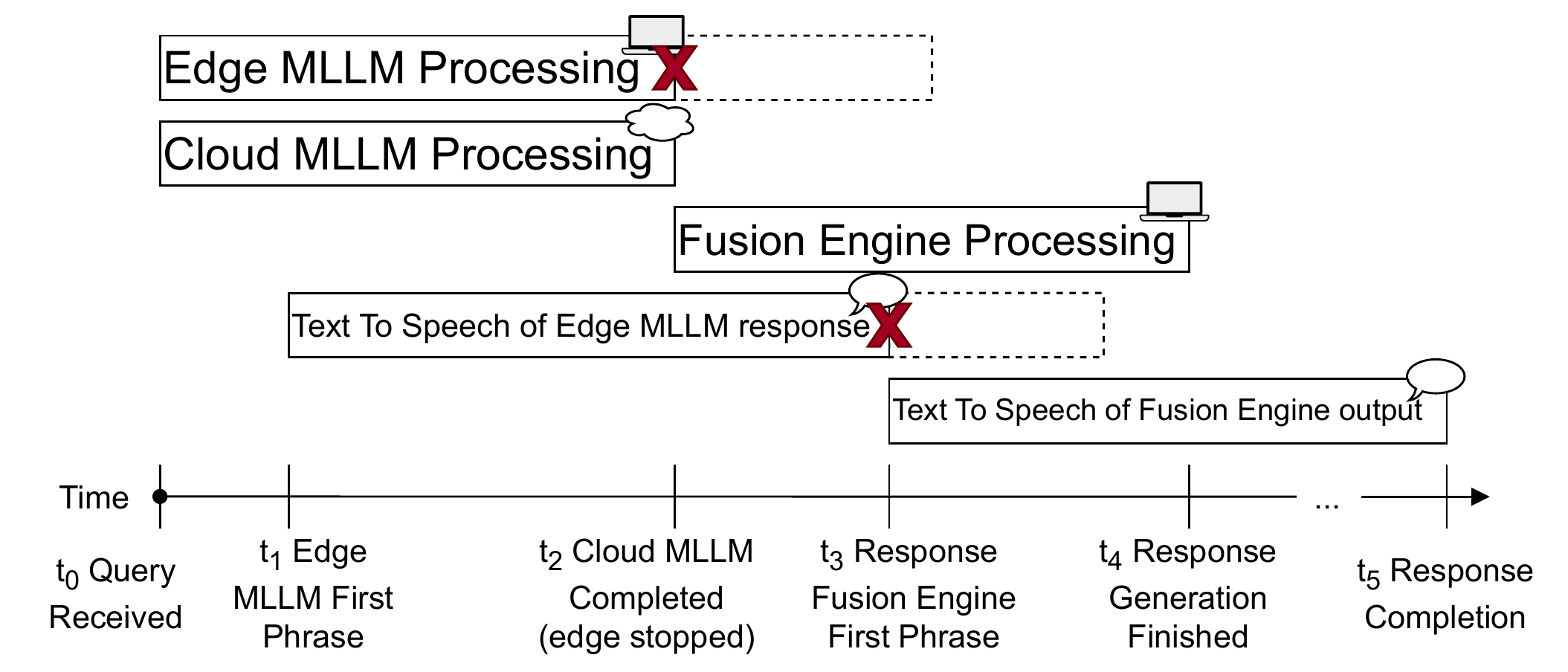}
    \caption{Schematic view of Response Fusion Engine. The edge MLLM processing is interrupted (red X symbol in the figure) once the higher quality cloud MLLM response is ready.}
    \Description{A timeline diagram in image_a69b7c.png showing the parallel processing sequence of Edge and Cloud MLLMs. At t0 (Query Received), both models begin processing simultaneously. Edge Text-To-Speech (TTS) starts early at t1. At t2, the Cloud MLLM completes its processing, which abruptly terminates the ongoing Edge MLLM and its TTS (both marked with red Xs) and immediately triggers the Fusion Engine Processing. The Fusion Engine begins its own TTS output at t3, finishes generating at t4, and the final response completes at t5.}
    \label{fig:FuseTimeDig}
\end{figure}

As described in Algorithm~\ref{Algorithm 1}, The Response Fusion Engine monitors the Cloud and Edge MLLMs and keeps track of which one begins generating first. If the cloud generates first, then the edge is not needed because it is assumed that the cloud is more accurate, so in this case, as shown in Line 8 of Algorithm~\ref{Algorithm 1}, the local is stopped and the cloud output is used instead. If the Edge begins streaming first, as shown in Line 14 and Figure~\ref{fig:FuseTimeDig}, the edge response is streamed to the TTS engine's queue ($Q_{\text{TTS}}$). Meanwhile, the cloud response is collected over time. When the full cloud response has been received, as denoted by {$t_2$} in Figure~\ref{fig:FuseTimeDig}, we prepare to run the Fusion Engine LLM. First, the edge MLLM is stopped to save resources, as shown in Line 17 of Algorithm~\ref{Algorithm 1}, and the necessary inputs for the Fusion Engine LLM are collected. The required inputs include the full cloud MLLM response and the local MLLM response up to the sentence or word being spoken at the time when the Fusion Engine LLM is estimated to stream its first phrase, denoted as {$t_3$} in Figure~\ref{fig:FuseTimeDig}. We define this estimated time value as the 95th Percentile of the first phrase arrival time ($0.5 s$) and denote it by $T_{est}$. This value, together with TTS speaking rate ($TTSR = 15$), is used to calculate the index of the last character, {$p$}, spoken by the TTS engine at the time when the Fusion Engine LLM begins streaming its response. The index is also limited (Line 19 to 27) to prevent mispronunciation of partial words.

The response data following this character is truncated from $Q_{\text{TTS}}$ in Line 28, and the Fusion Engine LLM is given the response generated by the edge MLLM up to character {$p$} along with the full cloud MLLM response. In the prompt, the Fusion Engine LLM is told to continue from the partial Edge LLM response and to treat the full Cloud MLLM response as the ground truth, correcting itself where necessary. It is also instructed to leave out redundant information.

\begin{algorithm}
\caption{Hybrid edge‑cloud MLLM output streaming with real-time fusion for seamless text-to-speech}
\label{Algorithm 1}
\begin{algorithmic}[1]

\State \textbf{Inputs:} $TTSR$ \Comment{TTS playback rate (characters/s)},\\ $T_{est}$ \Comment{Estimated 95th Percentile fusion first phrase time.}

\State \textbf{Initialize} two parallel MLLM workers:
\State\quad $R_\text{Edge}\gets$ \textbf{ConcurrentGenerate}(\textit{Q})
\State\quad $R_\text{Cloud}\gets$ \textbf{ConcurrentGenerate}(\textit{Q})

\State $t_0 \gets \text{CurrentTime}()$

\If{$\tau_{c1}\leq\tau_{e1}$}
    \State \textbf{Stop}(\textit{Edge MLLM})
    \State Append all \textit{Cloud MLLM} phrases to $Q_{\text{TTS}}$
    \State $t_2 \gets \text{CurrentTime}()$
\Else \Comment{first phrase arrives from \textit{Edge MLLM}}

    \State $t_1 \gets \text{CurrentTime}()$
    
    \While{Cloud MLLM Last Token hasn’t arrived}
        \State Append all \textit{Edge MLLM} phrases to $Q_{\text{TTS}}$
    \EndWhile

    \State $t_2 \gets \text{CurrentTime}()$

    \State \textbf{Stop}(\textit{Edge MLLM})

    \State $p \gets \lceil{TTSR\times(t_2 - t_1 + T_{est})}\rceil$ \Comment{predicted number of characters spoken}
    
    \If {$p > \text{length}(R_{\text{Edge}})$} 
    \State $p \gets \text{length}(R_{\text{Edge}})$ \Comment{Limit fusion Edge response input to Edge response length}
    \ElsIf{ $a \gets \text{earliest punctuation in } R_{\text{Edge}} \text{ from } p \text{ to end}$}
    \State $p \gets a$
    \ElsIf {$p$ \text{is in last word and end char is not punctuation}}
    \State $p \gets $ \text{Last character of previous word.} \Comment{Avoid partial words for accurate TTS pronunciation.}
    \Else
        \State $p \gets$ \text{Last character of current word.}
    \EndIf

    \State Truncate $Q_{\text{TTS}}$ after the first $p$ characters of $R_{\text{Edge}}$
    \State $R_{Fusion} \gets \text{StartFusionEngineLLM}\big(R_{\text{Edge}}[0\!:\!p],\,R_{\text{Cloud}}\big)$

    \While{Fusion Engine LLM is generating tokens} \Comment{This starts at $t_3$}
        \State Append all \textit{FE} phrases to $Q_{\text{TTS}}$
    \EndWhile
\EndIf

\end{algorithmic}
\end{algorithm}

A key element influencing the quality of an output is the seamlessness of audio generation, characterized by smooth transitions and uninterrupted audio feedback. This, in turn, depends on two additional factors: the rate of generating input text for the voice synthesizer from fusion engine and the speed of the generated audio, which is regulated by the utilized TTS model and user-selected speech rate. In this subsection, the factors influencing the text generation speed of the Response Fusion Engine are examined. These factors include external delays caused by interruptions in input streams from edge and cloud-based MLLMs, as well as internal delays resulting from latency within the internal LLM during processing and output generation process.

We calculate these factors from our experiment's chunked data using the following equations which correspond to Figure~\ref{fig:FuseTimeDig}. We iterate through phrases consisting of hard-punctuation-separated chunks, estimate their speaking time, and compare this to the time taken to generate each phrase from the LLM. Let $n_e$, $n_c$, and $n_m$ denote the number of phrases produced in the edge MLLM, cloud MLLM, and internal LLM inside Fusion Engine, respectively. Each phrase is associated with a corresponding character length: $L_e$, $L_c$, and $L_m$, representing the number of characters per phrase. Each phrase generation time is described by $\tau_e$, $\tau_c$, and $\tau_m$, corresponding to each phrase produced in the edge MLLM, cloud MLLM, and internal LLM, respectively.

The variable $t_0$ represents the start time of the system or interaction. The variable $t_1$ denotes the time at which the first edge MLLM phrase becomes available and is defined as:
\begin{equation}
t_1 = \tau_{e1}
\label{eq:eq1}
\end{equation}

The variable $t_2$ denotes the cloud turnaround time, i.e., the time at which the cloud response is completed and becomes available:
\begin{equation}
t_2 = \sum_{i=1}^{n_c}\tau_{ci}
\label{eq:eq2}
\end{equation}

The variable $t_3$ represents the time at which the LLM inside the Fusion Engine generates the first merged output phrase:

\begin{equation}
t_3 = t_2 + \tau_{m1}
\label{eq:eq3}
\end{equation}

The variable $t_4$ represents the time at which the last fusion phrase is received, indicating that the fusion process and the generation of the combined response have been completed:

\begin{equation}
t_4 = t_2 + \sum_{i=1}^{n_m}\tau_{mi}
\label{eq:eq4}
\end{equation}

As shown in Equations (\ref{eq:eq1})-(\ref{eq:eq4}), these time variables are influenced by the token generation rates of both edge and cloud MLLMs and the Fusion Engine LLM, and they vary depending on the input query.


The parameter $p$ represents the number of characters that must be read from the output of edge MLLM before the first merging phrase. This is estimated using $T_{est}$, the estimated 95th Percentile for the fusion LLM's first phrase time, and $t_1$ and $t_2$. The parameter $R$ represents the TTS speaking rate, expressed in characters per second. We use a buffer $b$ that increases based on how far ahead of the text to speech we are. If it drops below 0, we consider this an interruption.

The TTS module incorporates a buffer that receives input from the various components of the Response Management Module, as illustrated in Figure~\ref{fig:cog}. The parameter $C$ denotes the number of characters currently stored in the buffer. The state of this buffer serves as an indicator of interruptions during speech output, where a negative value of $C$ signifies that the TTS process has been interrupted. Such interruptions occur when the TTS speaking rate exceeds the rate at which tokens are streamed from components of the Response Management Module, resulting in temporary buffer underflow. In this section, to evaluate the seamlessness of the Fusion Engine generated response, we analyze the buffer state and measure the cumulative duration of TTS interruptions caused by the TTS outpacing Fusion Engine token streaming. This metric provides a quantitative assessment of the continuity and smoothness of the spoken output.

We derived mathematical formulas to quantify interruption metrics associated with the primary sources of interruption in the fusion system. For these calculations, we assume a constant TTS rate. Under this assumption, interruptions in speech synthesis (i.e., buffer underflow) can occur for two reasons: (1) the token generation rate is lower than the speaking rate, and (2) no tokens are generated while the TTS system is synthesizing speech. As mentioned earlier, the time interval associated with receiving each phrase from a model is used as the temporal basis of the system evaluation. The buffer state is evaluated at each such time step. Accordingly, upon the reception of the $k-th$ phrase from the model, the buffer state is updated according to Equation~\ref{eq:phrase}.

\begin{equation}
\label{eq:phrase}
\begin{array}{l}
\begin{aligned}
C_k &= 
\begin{cases}
|C_{k-1} + L_{k} - R \times \tau_{k}| & \text{if }C_{k-1} + L_{k} - R \times \tau_{k}i \ge 0 \\
0 & \text{otherwise}
\end{cases}
\end{aligned}
\end{array}
\end{equation}

The first source of interruption can be formalized based on the number of buffer-underflow occurrences during the token generation of each model that fills the buffer as shown in Equation \ref{eq:reason1},

\begin{equation}
U_n = \sum_{k=1}^{n} \mathbf{1}(C_{k-1} + L_{k} - R \times \tau_{k}i \le 0)
\label{eq:reason1}
\end{equation}

where $1(\cdot)$ is the indicator function, equal to 1 when the buffer occupancy is non-positive and 0 otherwise. The buffer-underflow causes a gap in speech synthesis, such that no character is available until the end of the next phrase time step. The total gap is computed as the sum of all intervals that follow a buffer-underflow state as shown in Equation \ref{eq:reason1time}, 

\begin{equation}
T_{\mathrm{gap}} = \sum_{k=1}^{n-1} \tau_{k+1}\,\mathbf{1}(C_{k-1} + L_{k} - R \times \tau_{k}i \le 0)
\label{eq:reason1time}
\end{equation}

Thus, whenever the buffer occupancy becomes non-positive, the subsequent time interval contributes to the gap. As previously mentioned, the second source of interruption arises when no tokens are generated during speech synthesis because the invoked model has not yet begun producing output tokens. Two key factors affect this metric: (i) the waiting time to receive the first token from the invoked model and (ii) the number of characters stored in the buffer ($C$) at the time of invocation. For example, $C_{FE\,LLM}$ denotes the number of characters stored in the buffer at the time the internal LLM within the Fusion Engine is invoked.

The first factor varies across the LLMs in our system. For the edge MLLM, it depends on the query and the edge computational resources. For the cloud MLLM, it depends on the query, network conditions, and other server-related parameters. For the Fusion Engine LLM, it depends on the edge computational resources, the input length, and the complexity of the fusion process.
The second factor, $C$, is equal to zero when the system invokes MLLMs (Algorithm 1, lines 4 and 5) but it can be nonzero ($C_{FE\,LLM}= TTSR \times T_{est}$) when the system invokes FE LLM. This factor plays a critical role in the interruption gap, as underestimating the FE LLM's first-phrase generation time in Algorithm 1 (line 18) can result in the truncation of excess buffered characters in line 28. Consequently, the interruption gap may increase. This gap can be quantified using Equation (8).

\begin{equation}
T_{\mathrm{FE\,gap}} = \tau_{m1} - T_{est}
\label{eq:BufferGap}
\end{equation}

According to Algorithm 1, if the condition in the \textit{if} statement (Line 7) is satisfied, the interruption is equal to the cloud MLLM first-token waiting time ($\tau_{c1}$) plus the total cloud MLLM gap, as defined in Equation~\ref{eq:reason1time}. Otherwise (Line 11), the interruption depends on $C_{FE\,LLM}$ and comprises the edge MLLM and Fusion Engine LLM first-token waiting times ($\tau_{e1}$ and $\tau_{m1}$), together with the total edge MLLM gap and the total local LLM gap, as defined in Equation \ref{eq:reason1time}.

The custom-defined metric $M$, as expressed in Equation \ref{eq:customM}, is designed to better reflect the impact of interruptions on the speech synthesis quality. It penalizes longer interruptions by scaling the interruption time exponentially. A larger M-value means a \textit{worse score} because there are more frequent or longer interruptions. Each interruption's time is increased by one before being raised to the power of 2 to ensure short interruptions are not effectively ignored.

\begin{equation} 
\begin{array}{l}
\begin{aligned}
M &= \sum_{i=1}^{n}\big(
\begin{cases}
(|C_{k-1} + L_{k} - R \times \tau_{k}i| + 1)^2 & \text{if }C_{k-1} + L_{k} - R \times \tau_{k}i < 0 \\
0 & \text{otherwise}
\end{cases} \big)
\end{aligned}
\end{array}
\label{eq:customM}
\end{equation}

\section{Evaluation}

\textbf{Experimental design.}
We tested our system's performance through both automated experiments with datasets, and through real-time interactions with actual blind and visually impaired experiment participants.

\subsection{Automated Experiments Setup}

\subsubsection{Dataset selection.}
To assess and enhance the system’s question-answering capabilities and contextual comprehension, we utilized the VizWiz dataset \cite{Gurari_2018_CVPR}, which consists of photographs captured by BLV individuals and provides a realistic benchmark for evaluating the system’s ability to address the real needs of BLV users through visual question answering tasks. For our system evaluation, we created a custom subset by selecting 200 images from the VizWiz dataset, with each image paired with two questions. To thoroughly evaluate the system’s performance across different domains, we categorized the image–question pairs into two groups: 100 pairs that are relevant to the Reasoning Engine’s expert AI Models, and other 100 pairs that are more general and unrelated to these experts and are instead processed by the Reasoning Engine’s generic AI Models. We further divided each group to explore different processing scenarios for both normal and urgent tasks. For half of the pairs, we generated questions in an urgent format, enabling the Urgency Detection module to identify and prioritize these tasks for faster processing via the fast-track pipeline. This approach allows for a comprehensive evaluation of the system’s adaptability to varying task urgency, as well as its ability to process pairs based on the queries' context.

\subsubsection{Hardware Setup.} 
The \name system, as depicted in Figure~\ref{fig:cog} and tested in the evaluation section, was implemented on a Windows 11 machine equipped with an Intel Core i9-10850K (10 cores / 20 threads), 32 GB of system RAM, and an NVIDIA RTX 3080 Ti with 12 GB of dedicated VRAM. This GPU served as the primary accelerator for the local MLLM inference with Gemma3-4b, object detection with yoloe-v8l-seg-pf \cite{cheng2024yolo}, and dense embedding extraction with the fine-tuned all-MiniLM-L6-v2. API calls were made on a 5G Wi-Fi connection in the region of North Texas. 

\subsection{System Performance Evaluation}
In this evaluation, we assessed the latency, accuracy, and supported speech rate of the proposed system. Through interviews with BLV users, we identified two primary factors affecting user satisfaction: system responsiveness(i.e., latency) and response quality.
The goal of designing the \name is to improve system responsiveness while maintaining high response quality. The \name’s Reasoning Engine employs multiple AI Models and MLLMs operating across the edge–cloud continuum. For urgent tasks, BLV users expect concise and immediate responses, whereas for non-urgent tasks, BLV users may prefer more detailed yet succinct outputs.
Response latency and quality for each task depend on how effectively the Cognition Module employs AI Models and MLLMs. Achieving this objective requires efficient task routing within the reasoning engine, which is enabled by accurate Urgency Detection and AI Router modules. An optimal solution would fully understand user queries and intents and route them to the most appropriate available AI engine.

\textbf{Latency Evaluation.} We measured two latency metrics: First Token Latency (TTFT) and End-to-End Latency (Turnaround Time). TTFT is defined as the time elapsed until the first word of the response is received, while Turnaround Time refers to the total time required to receive the complete response. BLV users typically employ different TTS applications with varying speech rates based on personal preference. Therefore, we excluded speech synthesis time and playback delays from our latency measurements to ensure a fair comparison.

\begin{figure}[t]
    \centering
    \includegraphics[clip,trim={0.2cm 0.4cm 0.2cm 0.2cm}, width=3.2in, alt={Two bar charts compare latency of Audo-Sight, Cloud-only (GPT-5.0), and Edge-only (gemma3:4b) systems on normal versus urgent queries. Chart (a) shows Audo-Sight's first token latency improving significantly from 5.5s on normal queries to 1.5s on urgent ones, approaching the sub-second speed of Edge-only, while Cloud-only remains the slowest at 12 to 14 seconds. Chart (b) shows end-to-end latencies remain relatively stable across query types, but notes that Audo-Sight shifts from making 0 edge model calls for normal queries to 127 calls for urgent ones.}]{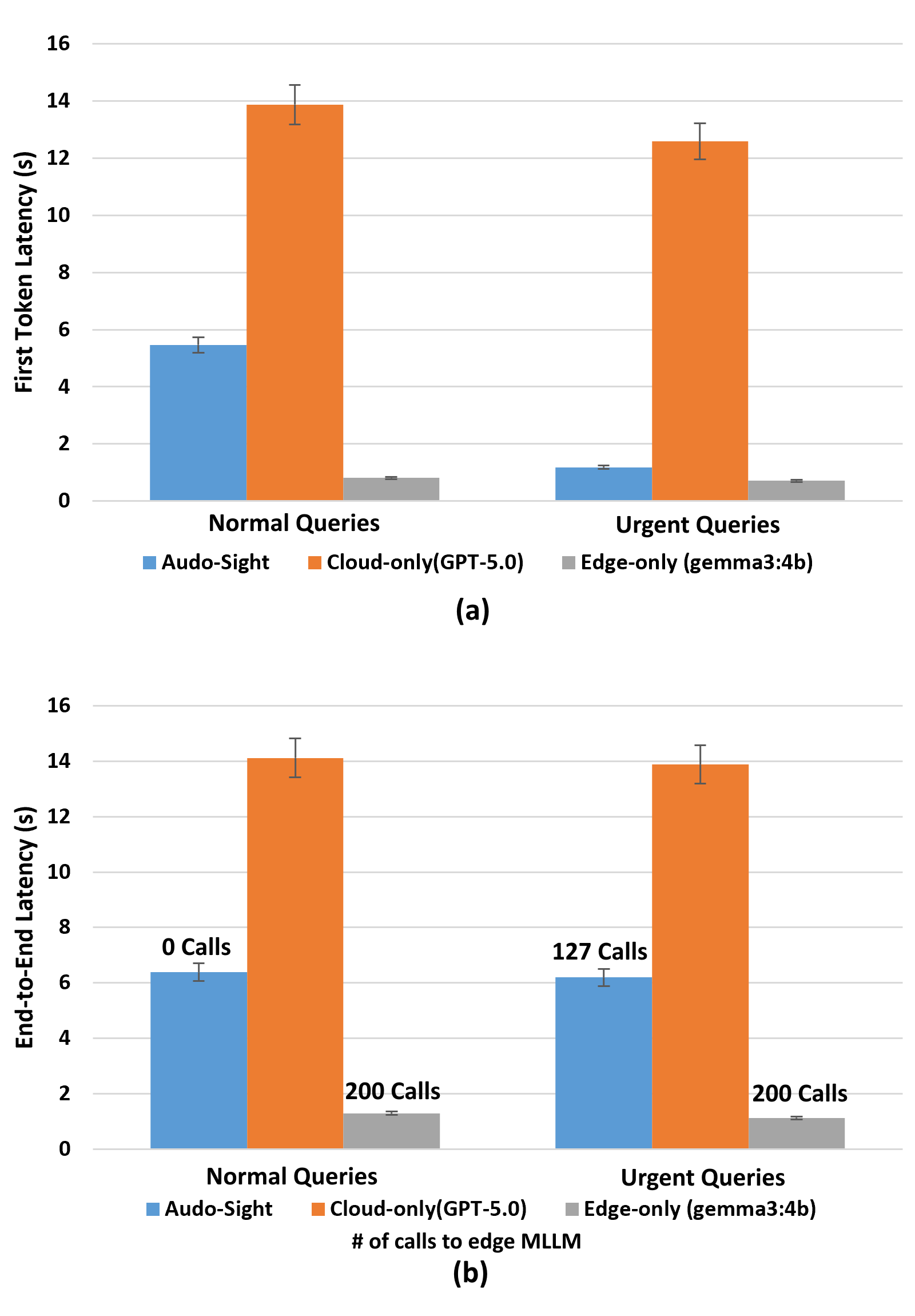}
    \caption{(a) First Token Latency (TTFT) comparison for urgent and normal tasks across three systems. (b) Comparison of end-to-end latency for urgent and normal tasks across three systems. Bars represent mean values obtained from experimental runs; error bars indicate 95\% confidence intervals.}
    \label{FigLatRes}
    \Description{Two bar charts compare latency of Audo-Sight, Cloud-only (GPT-5.0), and Edge-only (gemma3:4b) systems on normal versus urgent queries. Chart (a) shows Audo-Sight's first token latency improving significantly from 5.5s on normal queries to 1.5s on urgent ones, approaching the sub-second speed of Edge-only, while Cloud-only remains the slowest at 12 to 14 seconds. Chart (b) shows end-to-end latencies remain relatively stable across query types, but notes that Audo-Sight shifts from making 0 edge model calls for normal queries to 127 calls for urgent ones.}
\end{figure}

We present a comparison of TTFT and End-to-End Latency for both urgent and normal tasks across three systems: the proposed system \name, a local MLLM, and a cloud-based MLLM (GPT-5~\cite{openai_gpt5_2025}). As shown in the Figure~\ref{FigLatRes}(a), the proposed system significantly reduces TTFT for urgent tasks, achieving approximately 80\% improvements compared to the cloud-based MLLM. Similar performance gains are observed for normal tasks, where the proposed system improves TTFT by about 50\% relative to the cloud-based MLLM. For both urgent and normal tasks, the local MLLM demonstrates lower TTFT, achieving reductions of 71\% and 89\% relative to \name, respectively, and reductions of more than 94\% relative to cloud-based MLLMs. The relatively small errors in \name are expected, as it less frequently relies on cloud inference, which makes it less sensitive to network latency, cloud server load, and variability in response times. As a result, \name not only achieves lower latency overall, but also exhibits more predictable (low variance) response times.

The end-to-end latency of the proposed system is highly dependent on the end-to-end latency of the cloud-based MLLM. Considering the distribution of our experimental dataset, approximately half of the tasks are processed through the MLLM pipeline, while the remaining tasks are routed to expert AI Models in the \name system. For normal tasks not routed to expert AI models, \name relies on the cloud-based MLLM for response generation, resulting in no significant latency improvement compared to the cloud-based baseline. Similarly, for urgent tasks not routed to expert AI Models, the system also depends on the cloud-based MLLM to complete responses in the fusion engine. As shown in Figure~\ref{FigLatRes}(b), the average end-to-end latency of \name is roughly 50\% of that of the cloud-based MLLM, reflecting the partial routing of tasks to expert AI Models at the edge.
For both urgent and normal tasks, the edge-only system exhibits significantly lower end-to-end latency, achieving reductions of 80\% compared to \name. However, for normal queries, \name does not require any edge MLLM requests. In the case of urgent tasks, \name requires 40\% fewer edge calls compared to the edge-only system, leading to lower energy consumption on the edge. Additionally, as demonstrated in the next section, the accuracy results show that \name generates more accurate responses compared to the edge-only MLLMs.

Based on our observations, the difference between TTFT and end-to-end latency for tasks relying on cloud-based MLLMs is approximately 200 $ms$ on average and is negligible. This is because measuring the exact model-internal TTFT (i.e., the moment the first token is generated inside cloud-based MLLM servers) is impossible. Instead, we measure the earliest externally observable TTFT, which is heavily affected by network transmission, batching, framing, serialization, and other sources of latency. 

\begin{figure}[t]
    \centering
    \includegraphics[clip,trim={0.0cm 0cm 0.0cm 0cm}, width=3.3in, alt={A bar chart comparing the performance of two systems, Audo-Sight (blue bars) and Edge-only (gemma3:4b) (grey bars), across five evaluation metrics. The y-axis represents the "Avg. Score Across 4 Accuracy Metrics" on a scale from 0 to 100., Audo-Sight outperforms the Edge-only system in four out of the five categories: Correctness: Audo-Sight scores approximately 75, compared to Edge-only at roughly 64, Completeness: Audo-Sight scores approximately 78, compared to Edge-only at roughly 56, Faithfulness: Audo-Sight scores approximately 70, compared to Edge-only at roughly 56, Overall Acc.: Audo-Sight scores approximately 74, compared to Edge-only at roughly 62, The Edge-only system outperforms Audo-Sight in only one category, Clarity: Edge-only scores approximately 92, compared to Audo-Sight at roughly 78.}]{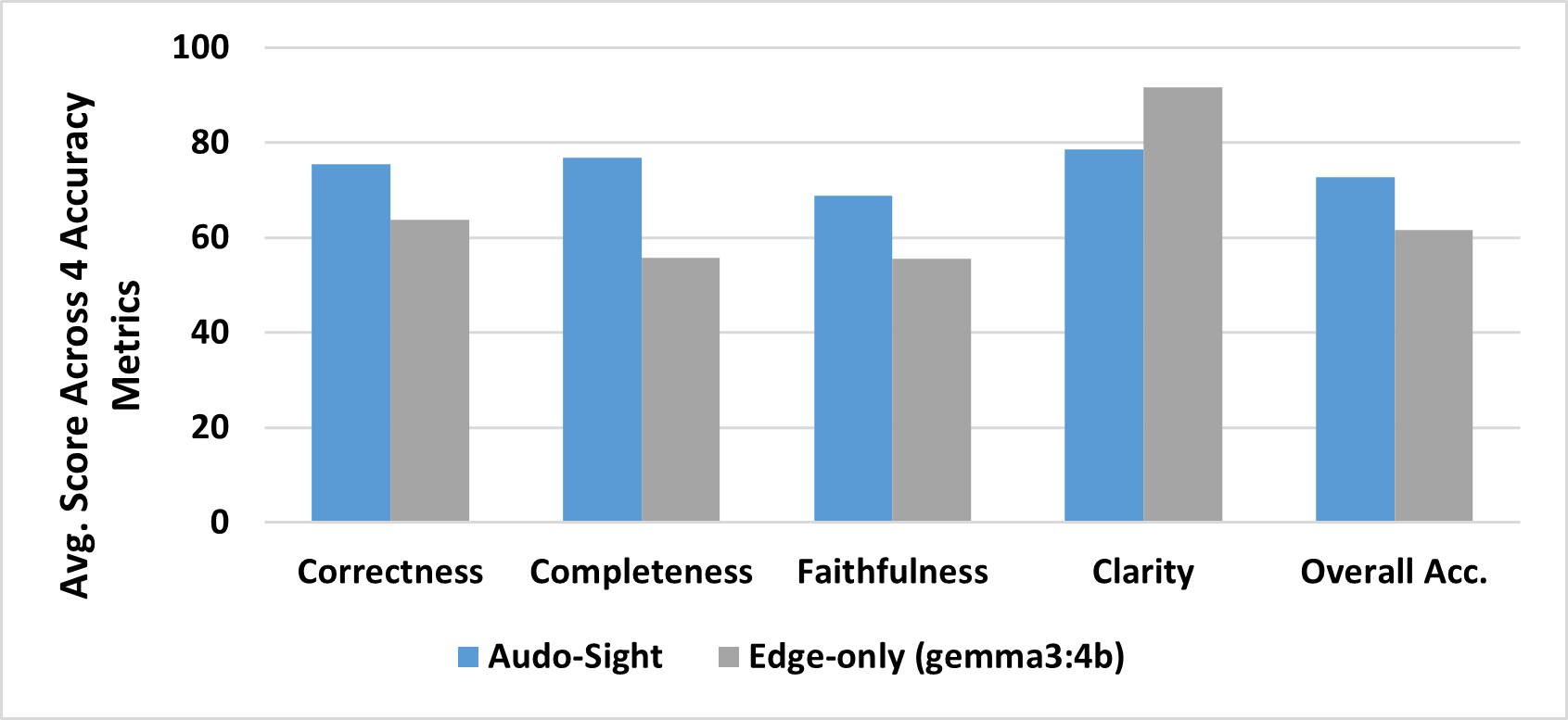}
    \caption{Accuracy comparison of \name and edge MLLMs responses across four metrics and their overall accuracy performance, with scores based on GPT-5 responses as the ground truth.}
    \label{FigScrRes}
    \Description{A bar chart comparing the performance of two systems, Audo-Sight (blue bars) and Edge-only (gemma3:4b) (grey bars), across five evaluation metrics. The y-axis represents the "Avg. Score Across 4 Accuracy Metrics" on a scale from 0 to 100., Audo-Sight outperforms the Edge-only system in four out of the five categories: Correctness: Audo-Sight scores approximately 75, compared to Edge-only at roughly 64, Completeness: Audo-Sight scores approximately 78, compared to Edge-only at roughly 56, Faithfulness: Audo-Sight scores approximately 70, compared to Edge-only at roughly 56, Overall Acc.: Audo-Sight scores approximately 74, compared to Edge-only at roughly 62, The Edge-only system outperforms Audo-Sight in only one category, Clarity: Edge-only scores approximately 92, compared to Audo-Sight at roughly 78.}
\end{figure}

\textbf{Accuracy Evaluation.} In question-answering systems, LLMs can be employed to assess the quality of responses generated by multiple models. In this study, we use LLMs as a scoring method to evaluate and compare the accuracy of three system-generated outputs. This approach requires a ground truth answer. Prior research ~\cite{BeMyAI,WorldScribe2024,Cheema:videodes,merchant:blvnav} has leveraged GPT as a robust MLLM in similar systems or studies. Consequently, we use our cloud-based MLLM, GPT-5, as the ground truth answer and evaluate how \name and our edge MLLMs, Gemma3:4b respond to a specific user query in comparison.
We selected GPT-5 as model to perform the evaluation and the process involves scoring each model’s response across four criteria: correctness (accuracy), completeness (extent to which key points from the ground truth are addressed), faithfulness (absence of incorrect or unsupported information), and clarity (organization and ease of understanding). Each criterion is rated on a scale from 1 to 5, and an overall score is derived, typically calculated as the average of the individual criteria ratings. Figure~\ref{FigScrRes} presents the scores for both \name and edge-based MLLMs. As shown, \name outperforms the Edge MLLMs in terms of correctness, completeness, faithfulness, and overall scores. However, Edge MLLMs achieve a higher score in the clarity criterion. This is because \name provides responses without additional refinements in cases of urgency, which can affect the clarity of the output.

\textbf{Speaking Rate Evaluation} The Max-TTS-Speed evaluation metrics is informed by BLV screen-reader usage and preferences from our own interviews and existing research. We noticed that many participants in our human study (discussed in section \ref{subsec:human}) use increased TTS speeds for screen-readers. Another study shows that many visually impaired individuals prefer high speaking rates, with some listening to over 500 words per minute, a speed that is 3X an average english speaking rate \cite{ListeningRates}. Therefore, the Max-TTS-Speed evaluation metric, denoted as $S_{max}$ reports the highest possible speaking rate of the system without interruption.
To determine $S_{max}$, we use use equation \ref{eq:lem1_sm_avg2}, which requires the set of phrases spoken by the system. Let $n$ represent the total number of phrases in a given response from the system, and let $\tau$ be the set of generation times per phrase. The first phrase generation time is considered to be zero since TTS output does not begin until after this phrase is generated. Let $L$ be the set containing the number of characters in each phrase chunk generated by the system in a given response. For each integer $i$ from 1 to $n$, we sum up the number of characters in the first $i$ phrases of the response. We then divide each of these cumulative character counts by the corresponding value of $\tau$. We call the resulting set $M$, which is the set of accumulated characters over their total time that has passed since text to speech started. By finding the minimum value in $M$, we obtain the highest sustainable character/second speaking rate that will avoid an interruption or gap in the text to speech output.

\begin{equation} \label{eq:lem1_sm_avg1}
\begin{array}{l}
\begin{aligned}
    M = \{ \frac{\sum_{k=1}^{i} L_k}{\tau_i} \mid i \in I\},
    \;\;\;\; I = \{1,2,\dots,n\}\\
\end{aligned} \\[0.5em]
\end{array}
\end{equation}

\begin{equation} \label{eq:lem1_sm_avg2}
\begin{array}{l}
\begin{aligned}
    S_\text{max} &= \min (M)
\end{aligned} \\[0.5em]
\end{array}
\end{equation}

Applying this formula to the system output across the various routes shown in fig. \ref{fig:cog}, the non-urgent generic route can always achieve up to 8.8X normal speaking rate or higher, and similarly, the non-urgent expert route can always achieve 10X normal speaking rate or higher. These rates are much higher than a typical preference for BLV people \cite{ListeningRates}. Since the full response is made available all at once in the route involving the expert models, the maximum speaking rate is always unlimited in this case. The case of urgent-generic involving the fusion engine will be discussed in-depth in section \ref{subsec:Merging}.

\subsection{Merging System Quality Evaluation} \label{subsec:Merging}
We evaluate the quality of the merging system using three criteria: Semantic Quality Evaluation, which assesses the accurate handling of redundant, contradictory, and supplemental information, Response Max Speed, and Seamlessness Evaluation.
\subsubsection{Speed and Seamlessness Evaluation}

\begin{figure}
    \centering
    \includegraphics[clip,trim={0.2cm 0.2cm 0.2cm 0.9cm}, width=3.3in, alt={A bar chart illustrating the relationship between Text-to-Speech (TTS) playback speed and response continuity. The vertical y-axis represents the percentage of "Responses without Interruption/Gap (\%)", ranging from 0 to 100 in increments of 10. The horizontal x-axis represents the "Supported TTS Speed (Factor of normal speech rate)", with categorical values of 0.25X, 0.5X, 1X, 1.25X, 1.5X, 2X, 2.5X, and 3X. The chart displays a single series of dark blue bars demonstrating a clear downward trend: as TTS speed increases, the percentage of uninterrupted responses steadily decreases. Approximate values for each bar are as follows: 0.25X: ~100\%, 0.5X: ~97\%, 1X: ~83\%, 1.25X: ~64\%, 1.5X: ~42\%, 2X:, 19\%, 2.5X: ~7\%,3X: ~3\%}]{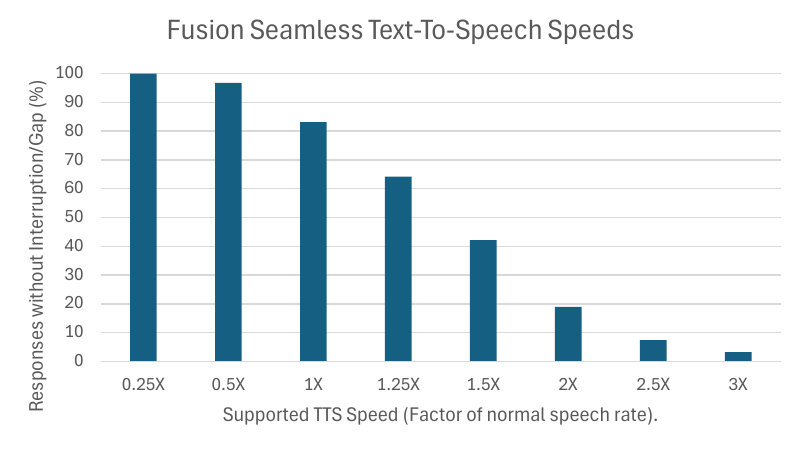}
    \caption{Percentage of Fusion responses without interruption at various Text-To-Speech Speeds (counting 15 characters/second as normal).}
    \label{fig:MaxTTSDistribution}
    \Description{A bar chart illustrating the relationship between Text-to-Speech (TTS) playback speed and response continuity. The vertical y-axis represents the percentage of "Responses without Interruption/Gap (\%)", ranging from 0 to 100 in increments of 10. The horizontal x-axis represents the "Supported TTS Speed (Factor of normal speech rate)", with categorical values of 0.25X, 0.5X, 1X, 1.25X, 1.5X, 2X, 2.5X, and 3X. The chart displays a single series of dark blue bars demonstrating a clear downward trend: as TTS speed increases, the percentage of uninterrupted responses steadily decreases. Approximate values for each bar are as follows: 0.25X: ~100\%, 0.5X: ~97\%, 1X: ~83\%, 1.25X: ~64\%, 1.5X: ~42\%, 2X:, 19\%, 2.5X: ~7\%,3X: ~3\%}
\end{figure}

\begin{figure}
    \centering
    \includegraphics[clip,trim={0.2cm 0.2cm 0.2cm 1.75cm}, width=3.3in, alt={Average Interruption time metric with Varied Voice Speed - Our tests show the system has minimal interruptions even when outputting at ~10x a normal speaking speed.}]{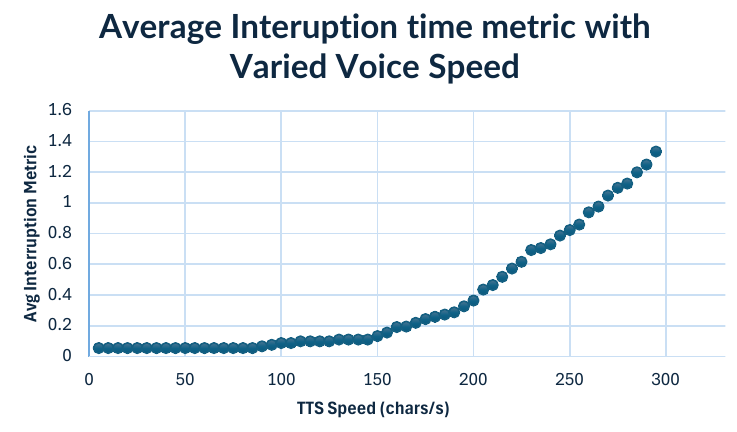}
    \caption{Average Interruption time metric with Varied Voice Speed - Our tests show the system has minimal interruptions even when outputting at ~10x a normal speaking speed.}
    \Description{Scatter plot depicting the relationship between Text-to-Speech (TTS) speed and the frequency of interruptions. X-axis: Labeled "TTS Speed (chars/s)", representing character output speed. The scale ranges from 0 to 300 in increments of 50. Y-axis: Labeled "Avg Interruption Metric", representing the average time or frequency of interruptions. The scale ranges from 0 to 1.6 in increments of 0.2. From 0 to approximately 150 chars/s, the data points form a relatively flat line, indicating that the average interruption metric remains very low (between 0.05 and 0.15). After 150 chars/s, the interruption metric begins to rise steadily. The upward trend shows the metric reaching approximately 0.4 at 200 chars/s, 0.8 at 250 chars/s, and peaking at approximately 1.4 at 300 chars/s.}
    \label{fig:Interruption Metric Graph}
\end{figure}

We ran a post hoc evaluation using our custom dataset with 100 urgent questions paired with images from the VizWiz dataset \cite{Gurari_2018_CVPR} and recorded the generation phrase timing to find $S_{max}$ for each response as defined in equations \ref{eq:lem1_sm_avg1} and \ref{eq:lem1_sm_avg2}. The results are shown in fig. \ref{fig:MaxTTSDistribution} and demonstrate a notable bottleneck compared to other routes taken by the system, which are discussed earlier in this section. When using the cloud or edge model in isolation, the difference in the response time of each model does not influence the maximum speed. However, when using the models in tandem, the difference in their response times must be masked by the text-to-speech output of the first model to ensure a seamless output. This principle is demonstrated by the second evaluation using the metric proposed in Equation \ref{eq:customM}. The same queries and response data are used to plot the graph in Fig. \ref{fig:Interruption Metric Graph}. This metric penalizes interruptions introduced by the merging system without including the "gap" between a complete edge MLLM response and the merged response. The results span many speaking rates, and demonstrate that interruptions are rare and short until around 150 characters per second, which is ten times an average speech rate. At a normal speaking rate, 84\% of responses have no interruption, and the rest have minimal interruption, but at a higher rates, interruptions become more common. One solution to this would be to create a "target speaking rate," where the text to speech speeds up or slows down dynamically based on expected cloud turnaround time and buffer size. Additionally, a smaller cloud model would increase the speaking, though most likely at the cost of some accuracy. For requests that are non-urgent and involve an expert model, the response is edited, and tests show that the text to speech rate supported is much faster than that of routes involving multi-modal LLM, which are the bottleneck of the system.


\subsubsection{Semantic Quality Evaluation.}
As discussed in Section~\ref{subsubsec:ResFusEng}, when an identical request is evaluated by two independent AI models, their respective responses may differ, potentially introducing redundant or conflicting information. A fusion system is considered effective when it delivers satisfactory outcomes across inputs that may contain redundant or conflicting information. To examine the performance, 
we first perform a semantic comparison of the two inputs for all cases processed by fusion engine module. Based on comparisons performed by three independent researchers, each input pair is classified into one of four categories. The first category, Equivalent (Eq), comprises pairs that contain identical information. The second category, Complementary Collaborative (CC), includes pairs that contribute additional information on the same topic (e.g., car color and car model). The third one, Distinct collaborative (DC), consists of pairs that provide correct information on different topics (e.g., in a scene description task, one input describes a car while the other describes a person within the same scene). The final category, Contradictory (Contra), includes pairs that contain conflicting information. The distribution of categories with the winning vote is as follows: Equivalent (48\%), Complementary Collaborative (23\%), Distinct Collaborative (2\%), and Contradictory (27\%).

To assess the semantic quality of our fusion engine outputs, we use responses generated by GPT-5 as a baseline. FLAN-T5\cite{chung2022scaling}, used in prior work \cite{jiang2023llmbln} as the GenFuser module and identified as an effective technique for integrating responses from multiple sources. It is included as a comparative method alongside our proposed approach. We compare the outputs of our fusion engine and those produced by FLAN-T5 against the baseline, using GPT-5 as a judge LLM. In this evaluation, both our fusion model and FLAN-T5 used identical prompts. FLAN-T5 allows for dynamic prompting, which allows us to evaluate its merging ability with our seamless response merging use case. For comparison, we also evaluate the quality of the concatenation merging strategy, that is simply forwarding the cloud output without any merging logic, effectively appending it to the local output spoken so far. 

To enable more effective evaluation, we define four metrics: Redundancy Handling (RH) is higher when the merged result contains less repetition of previously stated information. Contradictory Handling (CH) rewards correction where previously stated information contradicted the cloud response. Supplemental Handling (SH) is higher when new, complementary information is included in the merged response. Fluency is higher when the response is more grammatically correct and seamless. Claude Opus 4.7 was tasked with scoring all outputs according to these metrics, 30 percent of its judgments were reviewed by a human researcher and found to be largely accurate. Tables ~\ref{tab:qr_concat}, ~\ref{tab:qr_t5}, and ~\ref{tab:qr_gemma} present the scores for each category, along with the averages across all categories. FLAN-T5 achieves the lowest performance, with average scores in the 40-50 range. T5 performs worse than basic concatenation due to issues repeating previously spoken information. In contrast, our fusion engine module substantially outperforms both FLAN-T5 and concatenation overall, although it does not reach the performance level of the GPT-5 baseline and falls short of concatenation in certain situations. When judging how much supplemental information is included, concatenation performs better than Gemma3 model. This is likely because concatenation includes all the new information while the LLM tends to over-filter, which also results in a high redundancy handling score.

\begin{table}[h]
\centering
\caption{Merging Quality Across Response Relationships for Simple Concatenation (Concat).}
\label{tab:qr_concat}
\tagpdfsetup{table/header-rows=1}
\begin{tabular}{lccccc}
\toprule
\textbf{Metric} & \textbf{Eq} & \textbf{CC} & \textbf{DC} & \textbf{Contra} & \textbf{Avg} \\
\midrule
RH & 49.4 & 84.4 & 40.0 & 99.3 & 70.9 \\
CH & 62.6 & 60.9 & 90.0 & 67.4 & 64.0 \\
SH & 74.0 & 87.8 & 100.0 & 88.2 & 81.6 \\
Fluency & 35.3 & 43.5 & 40.0 & 50.4 & 41.4 \\
\midrule
All & 55.3 & 69.1 & 67.5 & 76.3 & 64.5 \\
\bottomrule
\end{tabular}
\end{table}

\begin{table}[h]
\centering
\caption{Merging Quality Across Response Relationships for Flan-T5.}
\label{tab:qr_t5}
\tagpdfsetup{table/header-rows=1}
\begin{tabular}{lccccc}
\toprule
\textbf{Metric} & \textbf{Eq} & \textbf{CC} & \textbf{DC} & \textbf{Contra} & \textbf{Avg} \\
\midrule
RH & 41.7 & 68.6 & 100.0 & 71.8 & 56.5 \\
CH & 47.0 & 37.1 & 30.0 & 40.9 & 42.9 \\
SH & 47.0 & 40.0 & 50.0 & 56.4 & 47.7 \\
Fluency & 34.8 & 43.8 & 60.0 & 38.2 & 38.2 \\
\midrule
All & 42.6 & 47.4 & 60.0 & 51.8 & 46.3 \\
\bottomrule
\end{tabular}
\end{table}

\begin{table}[h]
\centering
\caption{Merging Quality Across Response Relationships for Gemma3:4b (used by our system).}
\label{tab:qr_gemma}
\tagpdfsetup{table/header-rows=1}
\begin{tabular}{lccccc}
\toprule
\textbf{Metric} & \textbf{Eq} & \textbf{CC} & \textbf{DC} & \textbf{Contra} & \textbf{Avg} \\
\midrule
RH & 92.3 & 94.8 & 100.0 & 99.3 & 95.0 \\
CH & 62.6 & 53.9 & 60.0 & 68.2 & 62.0 \\
SH & 70.2 & 69.6 & 80.0 & 74.1 & 71.3 \\
Fluency & 71.9 & 66.1 & 50.0 & 75.6 & 71.1 \\
\midrule
All & 74.3 & 71.1 & 72.5 & 79.3 & 74.9 \\
\bottomrule
\end{tabular}
\end{table}

\subsection{User Evaluation (Human Study)}\label{subsec:human}

\textbf{Overview.} To test how the system performs for BVI individuals, we conducted an experiment with 8 legally blind participants, (aged 41-74, 6 female, 2 male; 6 with moderate or higher AI chat/voice assistant experience; 4 blind from birth). The experiment was performed virtually with pre-taken images from different environments and simulated situations to mitigate risks of real-world tests. The participants were asked to query the system as if they were in the given environment, and they provided feedback regarding the system's responses.

\textbf{Procedure.} Participants interacted with both our system and GPT 5 without knowing which one they were using. The participants entered their queries textually via a text prompt in our web app. 
The experiment was split into two situations, each of which involved three images in a sequence. This tested the system's ability to help the user complete multi-step goals, which included finding the restaurant counter and ordering food from a menu in the first situation, or meeting with a traveling companion at an airport and catching a flight in the second. Each situation was completed by each participant twice; once with \name and once with GPT-5 via OpenRouter. The system order was alternated for each participant to mitigate bias for or against the first system tested. After participants tested the system, the facilitating experimenter collected feedback from each participant regarding the system's design and trade-offs. The participants also offered potential improvements to the blind-friendliness Response Editor.

\textbf{Results.} When directly comparing the two systems, participants clearly demonstrate a preference for our system, with 62\% of responses saying they preferred guidance from \name overall compared to 15\% of responses preferring GPT-5. The users cited response expediency as a large reason in the direct comparisons and in ratings of each individual system, which correlates with our system's higher measured speed. Additionally, participants rated our system 15\% higher that GPT-5 in message quality, saying that our system's responses included a satisfactory amount of detail. When rating whether the system was sufficient for understanding the environment, on average, the participants rated the two systems to be equal in helping participants imagine the environment (4.1/5) and roughly equal in terms of clarity (\name - 4.1/5; GPT-5 - 4.3/5). In the discussion following the testing, participants praised the real-time merging system, saying that a fast, self-correcting system is better than a slow one in spite of higher initial accuracy, and they also voiced approval of the ability to control this feature by use of urgent language. The participants also offered some feedback regarding the quality of the responses which could be addressed in future work. The Blind-Friendly Response Editor could be extended to accommodate subjective preferences concerning phrases such as ``I see your point," or ``Look under your chair," or preferences for when to describe colors and visual details. Instructions such as ``Follow signs indicating XYZ," which rely on visual cues, can be avoided through a combination of the subtractive LLM-based Blind-Friendly Response Editor and MLLM-prompting. A suitable prompt combined with depth-sensors for distance information could potentially result in better scene descriptions in relation to the individual with specific angles and distances to provide clarity (e.g., it is positioned at 2 o'clock, 5 feet away). Additionally, participants suggested continuous updates would help maintain a user's trajectory during long, straight walks, which would be an improvement over the ``Walk straight" one-time instruction since any minor deviation will be amplified over significant distances.

\begin{figure}
    \centering
    \includegraphics[clip,trim={0.2cm 0.2cm 0.2cm 1.75cm}, width=3.3in, alt={BLV Participant Guidance Preference: 62\% of participants picked Audo-Sight when answering the question “Which system’s
guidance did you prefer overall?" 15\% picked chatgpt, and 23\% picked both performed equally.}]{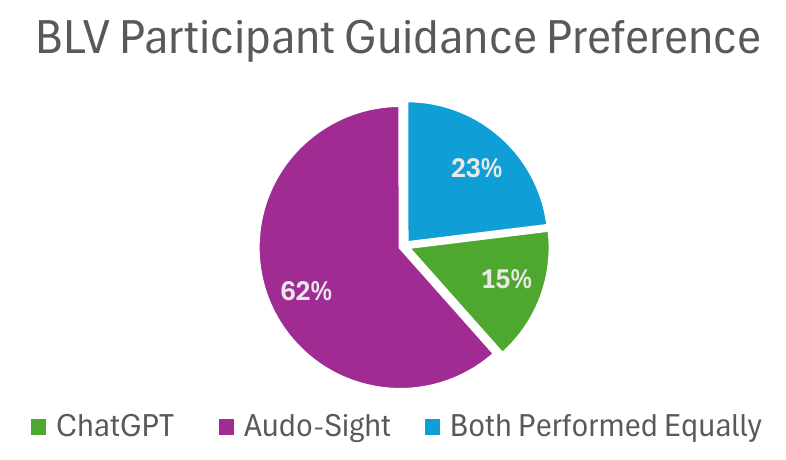}
    \caption{BLV Participant Guidance Preference: 62\% of responses were in favor of \name for the question ``Which system's guidance did you prefer overall?"}
    \Description{BLV Participant Guidance Preference: 62\% of participants picked Audo-Sight when answering the question “Which system’s
guidance did you prefer overall?" 15\% picked chatgpt, and 23\% picked both performed equally.}
    \label{fig:Participant Preference}
\end{figure}


\section{Discussion}

Although recent MLLM systems have demonstrated impressive performance, translating these advances into effective, user-centered assistive technologies is a significant challenge.
Despite the promising findings of this study to overcome this challenge, several remaining limitations should be acknowledged.

Importantly, the system performance evaluation was conducted using static images paired with user queries, and the current urgency detection mechanism relies primarily on the user's query. This setting does not fully capture the dynamic and continuous nature of real-world environments. Future work should therefore focus more on system performance evaluation methods that more closely resemble real-world usage patterns. For example, they can incorporate natural, multi-turn conversations in which an AI agent emulates a BLV user and continuously interacts with the system. In addition, short video sequences capturing evolving scenes could be used to evaluate the system performance during a context-aware assistance. Furthermore, incorporating visual information into the urgency detection and routing process could improve robustness and enable more accurate responses in rapidly changing or safety-critical situations. Nevertheless, in this study, we tried to mitigate these limitations by complementing the system performance evaluation with a human study that provided insights into how participants perceived the system during realistic and interactive scenarios. This allowed us to assess not only technical performance but also user experience, perceived usefulness, and practical considerations for real-world deployment. 

A key contribution of this study is the joint utilization of cloud-based and edge-based MLLMs and the essence of real-time fusion of their outputs. One may argue that with the rapid development of edge AI, such contribution will have short lifetime. We believe that even though both cloud and edge models continue to advance rapidly, fundamental differences between them are expected to persist due to their inherent characteristics. These differences include model capability and output quality, supported modalities, inference latency, user privacy, operational cost, energy consumption, and domain-specific knowledge, among other factors. In this work, we exploit the complementary strengths of both paradigms. Specifically, within the Response Management Module, we leverage the low latency of edge-based MLLM to provide rapid initial responses during urgent situations while compensating for their relatively lower accuracy by fusing these responses with those generated by a more capable cloud-based MLLM. In addition, we integrate Expert AI models, such as face recognition, for tasks that cannot be effectively or efficiently handled by general-purpose MLLMs. 
Consequently, despite the rapid evolution of AI models, the proposed framework, particularly its routing and response fusion mechanisms, will remain applicable in various systems (even beyond accessibility) in the foreseeable future. Rather than relying on the superiority of a single model, our approach unlocks effective \textit{integration} of multiple model with complementary capabilities to achieve a more robust and practical assistive system.

Many existing assistive systems rely heavily on cloud-based services, introducing challenges such as operational costs, privacy concerns, and the requirement for a continuous Internet connection. In contrast, the proposed framework reduces dependence on cloud services by establishing several functionalities to operate offline using edge-based models. Although this design may result in lower response quality in some scenarios, it offers advantages in terms of privacy, availability, reduced communication overhead, and resilience in environments with unreliable network connectivity. While this study focused on efficiency aspects, more exploration is needed to evaluate other aspects, such as privacy and network connectivity adaptation.

\section{Conclusion}
In this paper, we propose Audo-Sight, which represents a significant advancement in assistive technology for BLV individuals by integrating edge and cloud AI processing to provide timely, accurate, and context-aware conversational ambient perception. More technically, we sought to bridge the gap between the remarkable capabilities of MLLMs and their practical deployment challenges. 
The system's intelligent query routing, combined with multiple processing pipelines and the Response Fusion Engine, allows it to infer user intent effectively and generate responses quickly for urgent tasks while ensuring highly accurate and comprehensive outputs. Evaluations show that \name outperforms cloud-based MLLMs, starting responses to urgent tasks approximately 80\% faster and generating complete responses about 50\% faster across all tasks, with only a 25\% decrease in response quality with respect to high-end cloud MLLM. Furthermore, to improve robustness and availability, Audo-Sight's Reasoning Engine leverages both edge and cloud AI models. 

Our future direction is to improve the accuracy of the urgency detector to cover a wider range of implicitly urgent queries, as well as to enhance the AI Router's ability to better understand user intent. We also plan to make the Response Manager more intelligent by improving the blind-friendly Response Editor according to the experiences of interviewed BLV participants in this study. The Response Fusion Engine can also be improved by incorporating two-way communication between the Fusion Engine and the text-to-speech engine for better prediction of the spoken character count.
Another goal for our future design is to enhance the Reasoning Engine by incorporating additional expert AI models and reducing reliance on MLLMs through improved edge AI capabilities.

\section*{Acknowledgment}
We are thankful to the constructive comments of anonymous reviewers of the paper. We are also thankful to Aliyeh Haghollahi, Illango Zaruba, and Sai Donepudi who helped us in the results' verifications. This project is supported by NASA through MUREP ACEIR 2.0 Program,
Award\# 80NSSC25M0054 and NSF CAREER Award\#2419588.

\bibliographystyle{ACM-Reference-Format}
\bibliography{references}

\appendix
\section{Additional Human Testing Information}









\begin{figure}
    \centering
    \includegraphics[width=\columnwidth, clip,trim={0.0cm 0cm 0.0cm 0cm}, alt={A diverging stacked bar chart titled Audo-Sight Mid-Test Survey displaying Likert scale survey responses for five questions, Q1 through Q5. The chart aligns responses along a central Neutral baseline, with Agree and Strongly Agree stacking upwards, and Disagree and Strongly Disagree stacking downwards. Q1 shows mixed sentiment leaning positive with 46 percent agreement, 26 percent disagreement, and 26 percent neutral. Q2 leans heavily negative with 86 percent disagreement compared to 6 percent agreement and 12 percent neutral. Q3 and Q4 are entirely positive or neutral, with zero negative responses; Q3 shows 66 percent agreement, while Q4 shows 59 percent agreement. Q5 leans negative with 46 percent disagreement, 26 percent agreement, and 26 percent neutral responses.} width=3.3in]{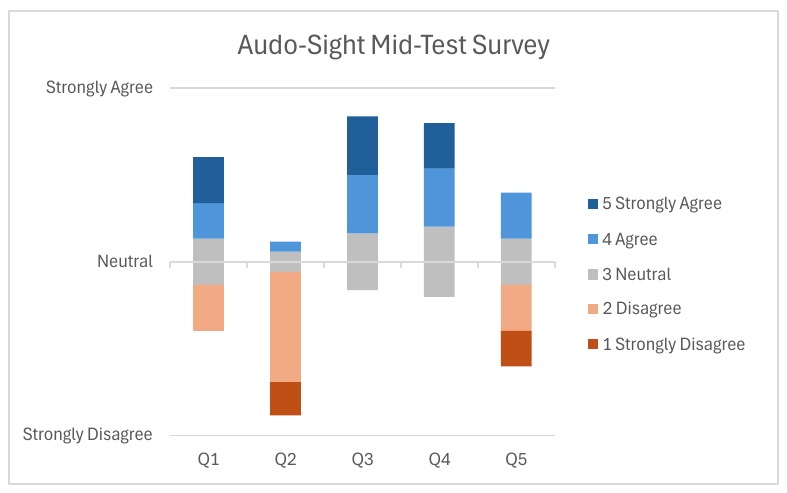}
    \caption{Human evaluation results after testing \name.}
    \label{Fig:DBC-AS}
    \Description{A diverging stacked bar chart titled Audo-Sight Mid-Test Survey displaying Likert scale survey responses for five questions, Q1 through Q5. The chart aligns responses along a central Neutral baseline, with Agree and Strongly Agree stacking upwards, and Disagree and Strongly Disagree stacking downwards. Q1 shows mixed sentiment leaning positive with 46 percent agreement, 26 percent disagreement, and 26 percent neutral. Q2 leans heavily negative with 86 percent disagreement compared to 6 percent agreement and 12 percent neutral. Q3 and Q4 are entirely positive or neutral, with zero negative responses; Q3 shows 66 percent agreement, while Q4 shows 59 percent agreement. Q5 leans negative with 46 percent disagreement, 26 percent agreement, and 26 percent neutral responses.}
\end{figure}

\begin{figure}[H]
    \centering
    \includegraphics[width=\columnwidth, clip,trim={0.0cm 0cm 0.0cm 0cm}, alt={A diverging stacked bar chart titled GPT-5 Mid-Test Survey displaying Likert scale survey responses for five questions, Q1 through Q5. The chart aligns responses along a central Neutral baseline, with Agree and Strongly Agree stacking upwards, and Disagree and Strongly Disagree stacking downwards. Q1 shows mostly positive sentiment with 53.9 percent agreement and 30.8 percent disagreement. Q2 leans negative with 61.6 percent disagreement compared to 15.4 percent agreement. Q3 and Q4 are entirely positive or neutral, showing high agreement at 92.3 percent and 84.6 percent respectively, with zero negative responses. Q5 is polarized with 53.9 percent disagreement, 46.2 percent agreement, and zero neutral responses.} width=3.3in]{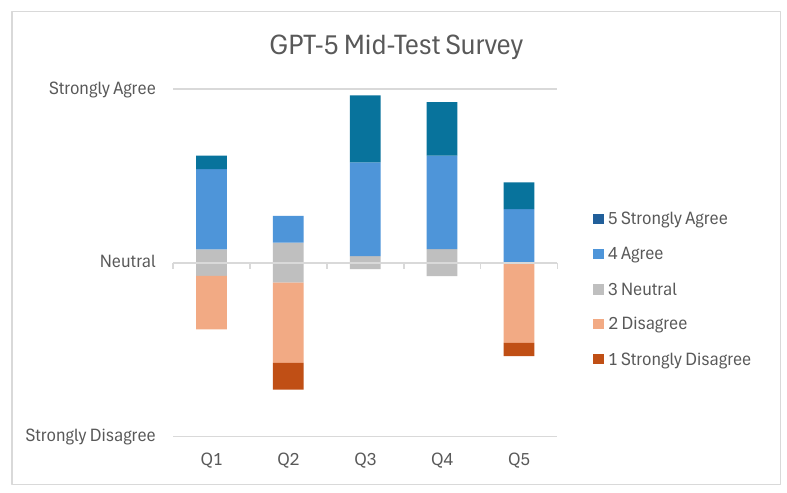}
    \caption{Human evaluation results after testing GPT-5.}
    \label{Fig:DBC-GPT}
    \Description{A diverging stacked bar chart titled GPT-5 Mid-Test Survey displaying Likert scale survey responses for five questions, Q1 through Q5. The chart aligns responses along a central Neutral baseline, with Agree and Strongly Agree stacking upwards, and Disagree and Strongly Disagree stacking downwards. Q1 shows mostly positive sentiment with 53.9 percent agreement and 30.8 percent disagreement. Q2 leans negative with 61.6 percent disagreement compared to 15.4 percent agreement. Q3 and Q4 are entirely positive or neutral, showing high agreement at 92.3 percent and 84.6 percent respectively, with zero negative responses. Q5 is polarized with 53.9 percent disagreement, 46.2 percent agreement, and zero neutral responses.}
\end{figure}

Tables \ref{tab:mid_test_questions} and \ref{tab:post_test_questions} list the questions given during the Human Study, and figures \ref{Fig:DBC-AS} and \ref{Fig:DBC-GPT} provide diverging bar charts showing the responses from the users. The bar charts' similar patterns may be explained by the structure of the testing procedure. The Mid-Test survey does not ask the users to compare the two systems with each-other, and the alternating order helps cancel out starting or ending bias for either system. \name outperforms GPT-5 in the post-test-survey when the users directly compare them after completing the full set of testing.

\begin{table*}[h]
\centering
\caption{Mid-Test Survey Questions}
\label{tab:mid_test_questions}
\tagpdfsetup{table/header-rows={1}}
\begin{tabular}{l | p{8.5cm} | p{4.5cm}}
\toprule
\textbf{ID} & \textbf{Question} & \textbf{Choices} \\
\midrule
Q1 & The system felt responsive or answered quickly. & 1-5 (Strongly agree, agree, neutral, disagree, strongly disagree) \\

Q2 & The information provided by the system felt too simple and left out too many details. & 1-5 (Strongly agree, agree, neutral, disagree, strongly disagree) \\

Q3 & The system’s explanations were clear and easy to understand. & 1-5 (Strongly agree, agree, neutral, disagree, strongly disagree) \\

Q4 & The system’s guidance helped me imagine and understand the surrounding environment. & 1-5 (Strongly agree, agree, neutral, disagree, strongly disagree) \\

Q5 & The system was unresponsive or took too long to answer. & 1-5 (Strongly agree, agree, neutral, disagree, strongly disagree) \\

Q6 & Were any of the system’s responses confusing or unhelpful in any way? If so, please explain what you found confusing or unhelpful and why. & Free response, optional \\

Q7 & Were any of the system’s responses offensive or insensitive? If so, please explain what you found offensive and why. & Free response, optional \\
\bottomrule
\end{tabular}
\end{table*}

\begin{table*}[h]
\centering
\caption{Post-Test Survey Questions}
\label{tab:post_test_questions}
\tagpdfsetup{table/header-rows={1}}
\begin{tabular}{l | p{8.5cm} | p{4.5cm}}
\toprule
\textbf{ID} & \textbf{Question} & \textbf{Choices} \\
\midrule
Q1P & Which AI guidance felt faster and more responsive? & System 1, system 2, both performed equally \\

Q2P & Which AI guidance were the most useful for achieving the urgent goal of catching your flight quickly? & System 1, system 2, both performed equally \\

Q3P & Overall, which AI guidance do you prefer? & System 1, system 2, both performed equally \\

Q4P & If you have any specific observations comparing the quality of the responses in the last two scenarios, please mention them here. & Free response, optional \\
\bottomrule
\end{tabular}
\end{table*}

\end{document}